\documentclass[twocolumn]{aastex631}
\usepackage[]{xcolor}
\usepackage{amsmath}
\usepackage{multirow}
\usepackage{hyperref}
\usepackage{textcomp}
\usepackage{cleveref}[2012/02/15]
\usepackage{fancyhdr}
\usepackage{graphicx}
\usepackage{amssymb}
\usepackage{longtable}
\usepackage{float}
\usepackage{dcolumn}
\usepackage{hhline}

\usepackage[toc,page]{appendix}
\usepackage[caption=false]{subfig}

\crefformat{footnote}{#2\footnotemark[#1]#3}

\usepackage{mathtools}
\makeatletter
\newcommand{\vast}{\bBigg@{5}}
\newcommand{\Vast}{\bBigg@{6}}
\makeatother

\newcommand{\kms}{km~s$^{-1}$}
\newcommand{\s}{$\sim$}
\newcommand{\n}{$-$}
\newcommand{\HI}{\ion{H}{1}}

\newcommand{\CIV}{\ion{C}{4}}

\newcommand{\SiIV}{\ion{Si}{4}}
\newcommand{\NV}{\ion{N}{5}}
\newcommand{\tm}{\tablenotemark} 
\newcommand{\tn}{\tablenotetext}
\newcommand{\RN}[1]{%
  \textup{\uppercase\expandafter{\romannumeral#1}}%
}
\defcitealias{Zheng_2019}{Z19}
\defcitealias{Wakker_2012}{W12}
\defcitealias{Savage_2009}{S09}

\shortauthors{Soto, C. et al.}

\begin{document}

\title{The Signature of the Northern Galactic Center Region in Low-Velocity UV Absorption}

\author[0000-0001-7840-2972]{Christian Soto}
\affiliation{Space Telescope Science Institute, 3700 San Martin Drive, Baltimore, MD 21218}
\email{csoto@stsci.edu}

\author[0000-0002-6541-869X]{Trisha Ashley}
\affiliation{Space Telescope Science Institute, 3700 San Martin Drive, Baltimore, MD 21218}
\email{tashley@stsci.edu}

\author[0000-0003-0724-4115]{Andrew J. Fox}
\affiliation{AURA for ESA, Space Telescope Science Institute, 3700 San Martin Drive, Baltimore, MD 21218}
\affiliation{Department of Physics \& Astronomy, Johns Hopkins University, 3400 N. Charles Street, Baltimore, MD 21218, USA}

\author[0000-0002-3120-7173]{Rongmon Bordoloi}
\affiliation{Department of Physics, North Carolina State University, 421 Riddick Hall, Raleigh, NC 27695-8202}

\begin{abstract}
The Galactic Center (GC) is surrounded by plasma lobes that extend up to \s14 kpc above and below the plane. 
Until now, UV absorption studies of these lobes have only focused on high-velocity components ($|v_{\rm LSR}|>100$ \kms) because low- and intermediate-velocity (LIV) components ($|v_{\rm LSR}|<100$ \kms) are blended with 
foreground interstellar medium. To overcome this difficulty, we present a differential experiment to compare the LIV absorption between different structures within the GC region, including the Fermi Bubbles (FBs; seen in $\gamma$-rays), the eROSITA Bubbles (eBs; seen in X-rays), and the Loop I North Polar Spur (LNPS) association, an X-ray and radio feature within the northern eB. 
We use far-UV spectra from Hubble Space Telescope to measure 
LIV \SiIV\ absorption 
in 61 AGN sight lines, of which 21 pass through the FBs, 53 pass through the 
eBs, and 18 pass through the LNPS. We also compare our measurements to those in the literature from sight lines covering the disk-halo interface and CGM. 
We find that the FBs and eBs have enhancements in measured columns of 
0.22-0.29 dex in log. We also remove the contribution of a modeled disk and 
CGM component from the measured \SiIV\ columns and find that the northern eB still retains a \SiIV\ enhancement of 
0.62 dex in log. A similar enhancement is not seen in the southern eB.  Since 
the LNPS model-subtracted residuals show an enhancement compared to the rest of the northern eB of 0.69 dex, the northern eB enhancement may be caused by the LNPS. 

\end{abstract}


\section{Introduction}\label{section:intro}

The Fermi Bubbles (FBs) are two bipolar plasma lobes launched from the center of the Milky Way, reaching \s55\degr\ in Galactic latitude 
above and below the Galactic plane. Given their close proximity, we can use them as a laboratory to study the effect of nuclear feedback on the baryonic ecosystems of galaxies in greater detail than is possible in any other galaxy. Recent evidence suggests that the FBs likely formed $\approx$3-6 Myr ago through an energetic outburst from Sagittarius~A* \citep{Zubovas_2011, Guo_2012, Yang_2012, Bland_Hawthorn_2013, Bland_Hawthorn_2019,  Mou_2014, Fox_2020, Yang_2022}. For example, \citet{Bland_Hawthorn_2013, Bland_Hawthorn_2019} and \citet{Fox_2020} find an excess in H$\alpha$, \CIV, and \SiIV\ in the Magellanic Stream passing directly below the Galactic Center (GC). Models produced by \citet{Bland_Hawthorn_2019} show that a Seyfert flare coming from Sagittarius~A* $3.5\pm1$ Myr ago is necessary to produce those elevated present-day levels of H$\alpha$ in the Magellanic Stream. This timescale agrees with models that suggest the FBs formed via a jet or flare emanating from Sagittarius~A* in the past 1-6 Myr \citep{Zubovas_2011, Guo_2012, Yang_2012}. 
The timescale also agrees with the 6-9 Myr age of the UV-observed outflow modeled by \citet{Bordoloi_2017}. 
 An alternative hypothesis for the growth of the FBs is nuclear star formation; although this would operate on a longer timescale of $30-100$ Myr \citep{Yusef_Zadeh_2009, Lacki_2014, Crocker_2015, Sarkar_2015}.  

While the FBs are defined by the gamma-ray emitting regions of the nuclear outflow, counterparts at other wavelengths have also been detected in microwave, optical, and polarized radio emission \citep{Bland_Hawthorn_2003, Dobler_2008, Su_2010, Dobler_2010, Carretti_2013, Ackermann_2014,  Krishnarao_2020}.  Recently, an X-ray counterpart has also been discovered, the eROSITA Bubbles \citep[eBs; ][]{Predehl_2020}. The eBs are similar in shape but extend further than the FBs, reaching $\pm85\degr$ in latitude or \s14 kpc in height. The northern eB also encompasses the multi-wavelength feature, called the Loop I and North Polar Spur (hereafter, LNPS) association \citep{Hanbury_1960, Bowyer_1968, Berkhuijsen_1971, Haslam_1982, Sofue_1979, Sun_2013, PlanckCollaboration_2016, Predehl_2020}. The origin of the LNPS has long been debated. It is believed to trace either nearby supernova remnant projected in front of the GC \s$100-200$ pc away or a structure physically associated with the GC, or an overlapping projection of both a GC feature and nearby supernova remnant \citep[see][and references therein]{Lallement_2022}. Until recently, one strong argument against the LNPS being associated with the GC was that the LNPS had no southern Galactic counterpart. The discovery of a southern X-ray bubble (a.k.a. the southern eB) has reignited that debate \citep[][]{Predehl_2020, Yang_2022, Lallement_2022}. 

Discrete high-velocity clouds (HVCs) thought to be associated with the FBs have been detected in \HI\ emission, CO emission, UV H$_2$ absorption, and UV-metal absorption towards AGN sight lines \citep{McClure_Griffiths_2013, Fox_2015, Bordoloi_2017, Savage_2017, Di_Teodoro_2018, Di_Teodoro_2020, Karim_2018, Lockman_2020, Ashley_2020, Ashley_2022, Cashman_2021}. The velocities of these clouds ($|v_{\rm LSR}|>100$ \kms) are high enough to distinguish them from Milky Way disk gas at $|v_{\rm LSR}|<100$ \kms. On the other hand, diffuse gas at \textit{low} and \textit{intermediate} velocities (LIV; $|v_{\rm LSR}| <100$ \kms) have largely gone unexplored  because they overlap with the Milky Way interstellar medium in velocity space, and are thus contaminated by foreground absorption.
 
This contamination issue can be overcome with a differential experiment in which absorption in FB sight lines is compared to absorption in nearby sight lines outside the bubbles. Since the plasma inside the FBs has a high ionization level, sight lines passing inside the bubbles are expected to show stronger high-ion absorption than sight lines passing outside.  High ions are therefore an ideal tracer for this experiment. For example, \citet{Savage_2017} study a pair foreground-background stellar sightlines, which show enhancement of LIV \SiIV\ and \CIV\ absorption towards the southern FB, but they do not explore its significance. This effect was also seen by \citet[][hereafter W12]{Wakker_2012} and \citet[][hereafter Z19]{Zheng_2019}, who measured the amount of LIV \SiIV\ absorption towards 58 and 132 AGN sight lines, respectively, through the disk-halo interface and CGM, probing gas at temperatures of \s$10^{4}-10^{5}$K \citep{Gnat_2007}. \citetalias{Wakker_2012} and \citetalias{Zheng_2019} find an excess of 0.1--0.26 dex in LIV \SiIV\ column densities towards the GC. \citetalias{Zheng_2019} suggest that this enhancement could be due to the FBs, but they do not explore it further. 

In this paper, we present a differential survey of LIV \SiIV\ absorption in sight lines that pass through various structures in the GC region, including the FBs, eBs, and LNPS, using UV spectra from the Hubble Space Telescope Cosmic Origins Spectrograph (HST/COS) and  Space Telescope Imaging Spectrograph (HST/STIS).  We chose to use \SiIV\ doublet for this study because among the high-ion doublets in the COS bandpass, \NV\ is often undetected and \CIV\ is often saturated, while \SiIV\ tends to yield a measurable column density \citep{Fox_2015, Bordoloi_2017, Karim_2018, Ashley_2020}.  Additionally, using \SiIV\ allows us to make direct comparisons of GC pointings to the extensive sample in \citet[][hereafter S09]{Savage_2009}, \citetalias{Wakker_2012}, and \citetalias{Zheng_2019} probing the rest of the disk-halo interface and CGM.  In Section~\ref{section:observations} we present our data-set and discuss our observations and data reduction. In Section~\ref{section:results} we present the results of our survey and discuss how our results compare to those in the literature. Our conclusions are then presented in Section~\ref{section:conclusion}.

\section{Observations and Data Reduction}\label{section:observations} 

\begin{figure*}[!ht]
    \centering
    \epsscale{1.2} 
    \plotone{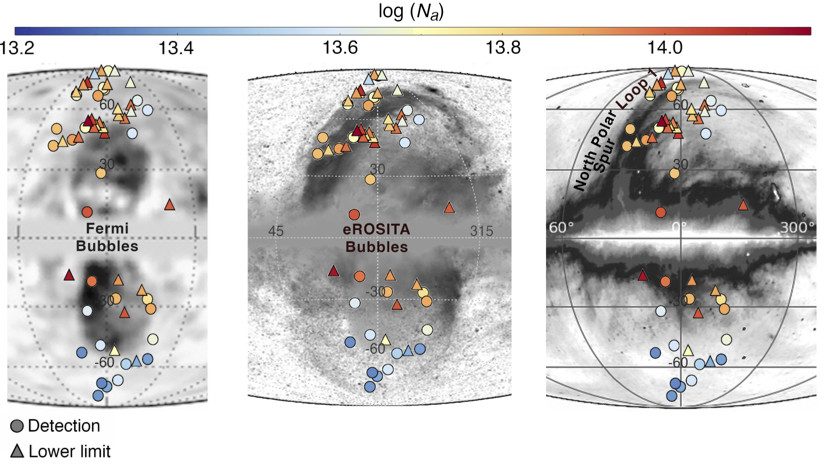}
\caption {Maps showing the location of our sample sight lines and their log $N_{a}$(\SiIV) values ($N_{a}$ in cm$^{-2}$) denoted by the colorscale. 
The circles are background quasars and the triangles indicate lower limits due to saturation. The background greyscales  display the extent of the FBs, eBs, and LNPS association using the following adapted maps: (Left) a gamma-ray map from \citet{Ackermann_2014} in Mollweide projection, (Middle) a combined gamma-ray and X-ray map from \citet{Predehl_2020} in Hammer-Aitoff projection, and (Right) a 408 MHz emission map from \citet{Haslam_1982} and reprocessed by \citet{Remazeilles_2015} in Mollweide projection.
\label{figure:logN}}
\end{figure*}

Our sample consists of 61 AGN sight lines with HST/COS and HST/STIS data drawn from the samples of \citet{Fox_2015}, \citet{Bordoloi_2017}, \citet{Karim_2018}, and \citet{Ashley_2020}; 21 pass through the FBs, 53 pass through the eBs, 18 pass through the LNPS association, and 8 pass through none of these GC features. 
For the sight lines passing through the FBs, the set also included targets that were listed in the literature as being on the ``edge" of the bubble. Seven of the sight lines in our sample are close to the Magellanic System or have HVCs associated with the Magellanic Stream identified in their spectrum \cite[see Table~\ref{table:Soto};][]{Fox_2014, Karim_2018}. As discussed in Section~\ref{section:results}, the Magellanic Stream does not contribute an excess of LIV \SiIV\ absorption to our measurements. Figure~\ref{figure:logN} shows the location of the sight lines relative to the FBs, eBs, and LNPS. In Table~\ref{table:Soto}, we present basic information on each sight line. 

Each sight line has a COS FUV G130M spectrum and one sight line (NGC5548) has STIS E140M spectrum. The COS data were initially calibrated with the \texttt{calcos} calibration pipeline. We then applied customized velocity alignment and co-addition steps described in \citet{Wakker_2015}, which align the Galactic absorption with \HI\ emission. 

\begin{figure}[!hb]
    \centering
    \epsscale{1.2} 
    \plotone{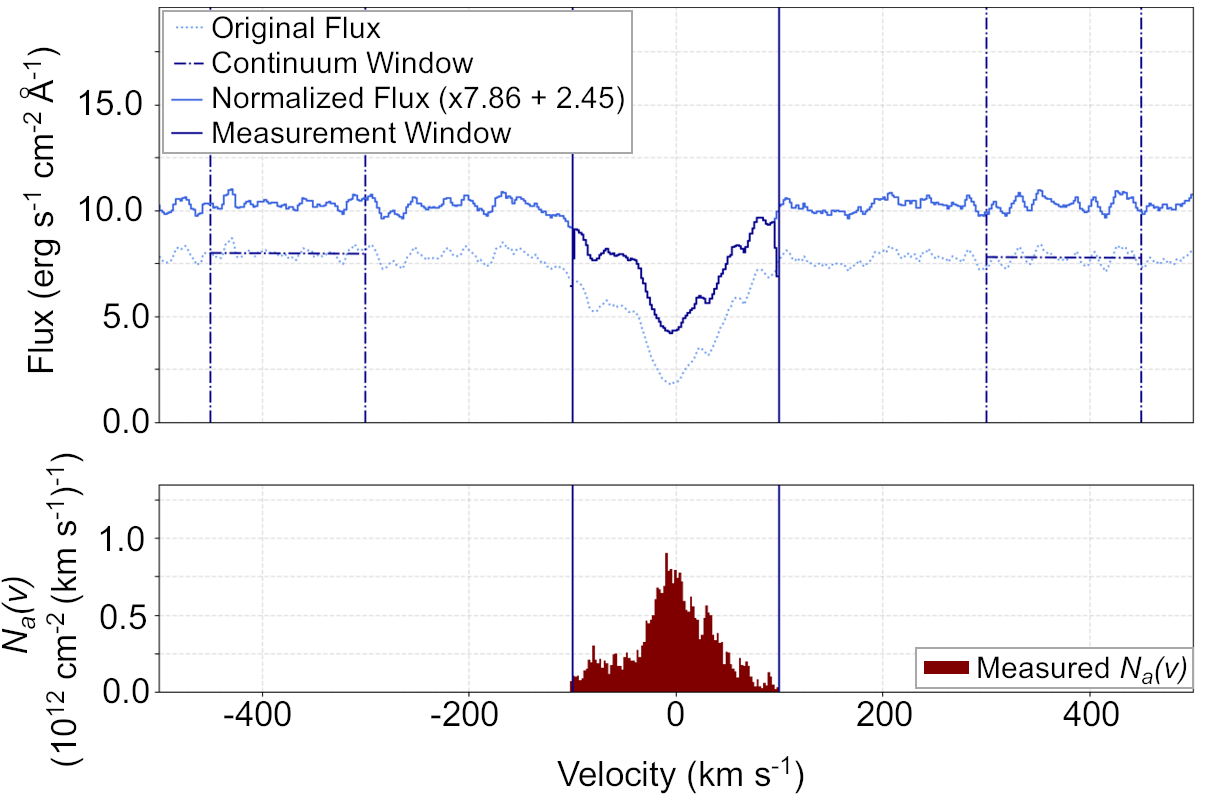}
\caption {Example spectral plots from the \texttt{spectrAOD} code for the sight line toward QSO ESO\,141-G55, which passes through the southern FB. The top panel shows the \SiIV\ 1393 line profile. The lower panel shows the ACD as a function of LSR velocity. 
\label{figure:spectrAOD}}
\end{figure}

We measure the apparent optical depths (AODs) and the apparent column densities (ACDs) of the LIV \SiIV\  absorption using a custom Python package called \texttt{spectrAOD}\footnote{https://github.com/cmagness/spectrAOD} \citep{Magness_2020}. AODs are used to analyze interstellar absorption lines; they represent the true optical depth with the addition of instrumental blurring \citep{Savage_1991}. To make these measurements, \texttt{spectraAOD} first normalizes the spectrum using a straight line between two sections of user-determined continuum.  Continuum ranges were chosen for each ion visually, avoiding absorption features such as redshifted intergalactic absorbers and HVCs identified in previous work \citep{Fox_2015, Bordoloi_2017, Karim_2018}. The velocity-dependent AOD, $\tau_{a}(v)$, is calculated as:

\begin{equation}
\tau_{a}(v)= -\mathrm{ln}[F_{c}(v)/F(v)],
\end{equation}

\noindent where $F(v)$ is the flux of the absorption and $F_{c}(v)$ is the continuum flux. $\tau_{a}(v)$ is then used to find the apparent column density ($N_{a}$):

\begin{equation}
N_{a} = \int_{v_-}^{v_+} 3.768\times10^{14}\  \dfrac{\tau_{a}(v)}{f\lambda} dv~\mathrm{cm}^{-2},
\end{equation}

\noindent where $v_-$ and $v_+$ are the LSR velocity limits of the absorption, $f$ is oscillator strength of the transition of interest \citep{Morton_2003}, and $\lambda$ is the transition wavelength in Angstroms. For each sight line, we measure the AOD and ACD between $-100$ and $100$ \kms\ for \SiIV\ $\lambda\lambda$1393 and 1402. Figure~\ref{figure:spectrAOD} shows an example of \texttt{spectraAOD}'s graphical output, including the sight lines' spectra, continuum fits, and ACD measurements. 

\texttt{spectrAOD} determines the noise using the standard deviation of the flux around the continuum. If the absorption reaches a minimum flux less than the noise anywhere within the line profile, then it is labeled as saturated. If the minimum flux is greater than the noise and the line is detected at 3$\sigma$ significance (i.e., the equivalent width is at least three times the equivalent width error), then the absorption is labeled as detected. In all other cases, the absorption is labeled as not detected. 

For lines that are unblended and unsaturated, the measured ACD from each member of the doublet should match within their respective measurement errors. However, different measurements may arise due to contamination from intergalactic absorption, continuum-fit errors, and/or unresolved saturation. The column-density measurements from each member of the doublet were considered to match if the following condition was met: 

\begin{equation}\label{equation:X-value}
\dfrac{{\rm log}(N_{a2})-{\rm log}(N_{a1})}{\sqrt{\sigma_{2}^2+\sigma_{1}^2}}< 3
\end{equation}

\noindent where $\sigma$ is the error on each logarithmic ACD measurement. If the value was $>$3, then we inspected the spectrum to determine which doublet line was more reliable. For example, if an intergalactic absorber was found in the stronger line, then the weaker lines measurement was considered more reliable and was used. 

In cases where both members of the \SiIV\ doublet were detected and unsaturated, and the requirement of Equation~\ref{equation:X-value} was met, we used the average of the two ACD measurements in our analysis. If both lines were saturated and visual inspection of the data did not reveal any issues with either line, then the weaker line was used to derive a lower limit on the ACD. If the stronger line was saturated and the weaker line was detected without apparent saturation, then we adopt the measurement from the weaker line. In this case, the weaker line was labeled as detected and unsaturated if the Equation~\ref{equation:X-value} requirement was met, otherwise it was labeled as saturated to account for unresolved saturation.  

\section{Results and Discussion}\label{section:results} 

Of the 61 sight lines measured in our study, one sight line (SDSS J141038.40+230447.0) was unusable due to intergalactic absorption blending in both \SiIV\ lines. Of the remaining 60 sight lines, the \SiIV\ detection rate was 100\%. For sight lines passing though the FBs, eBs, and LNPS 45\%, 39\%, and 60\%\ respectively, are labeled as saturated.


\begin{figure*}[!ht]
    \centering
    \epsscale{1.1} 
\plotone{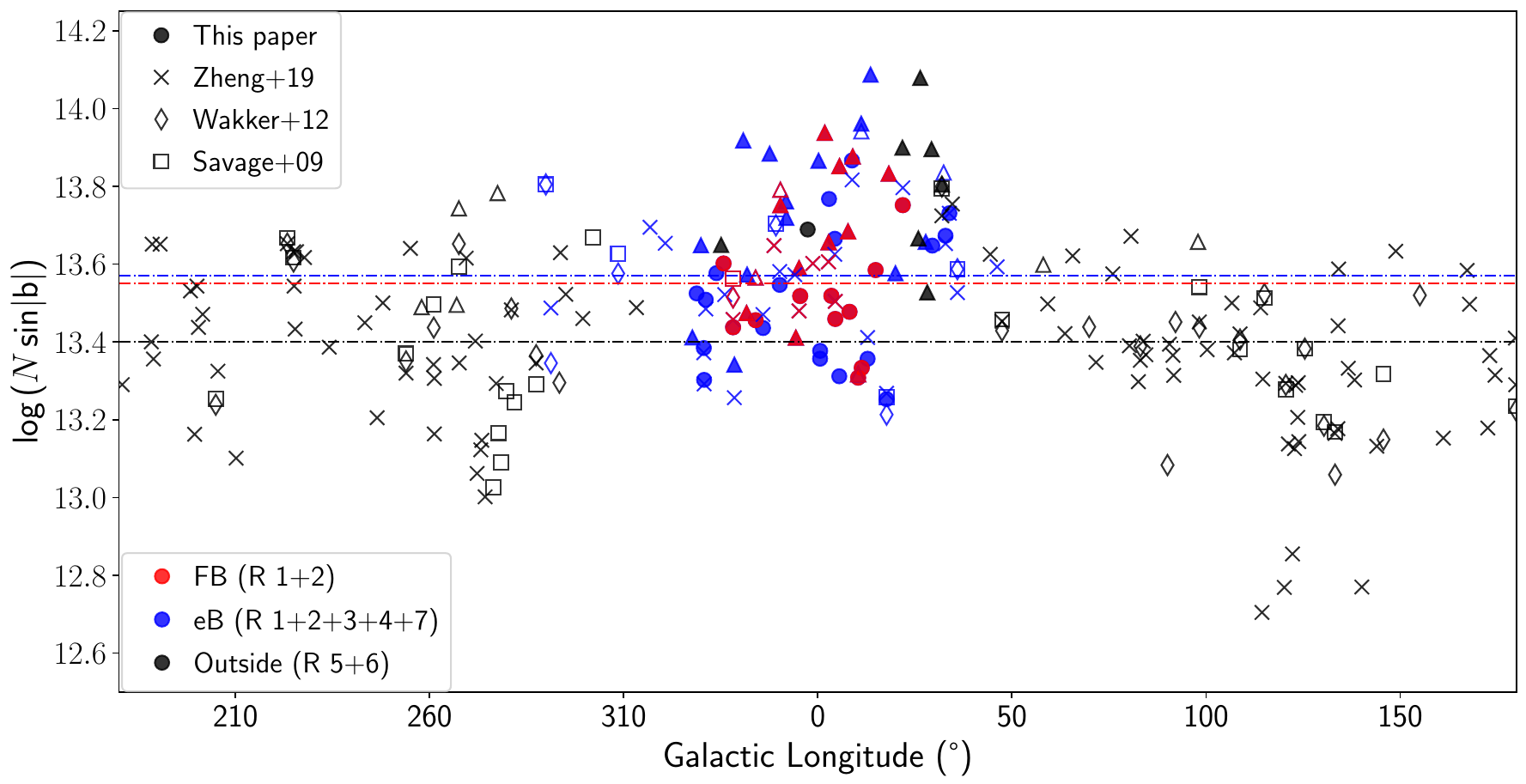}\\  \plotone{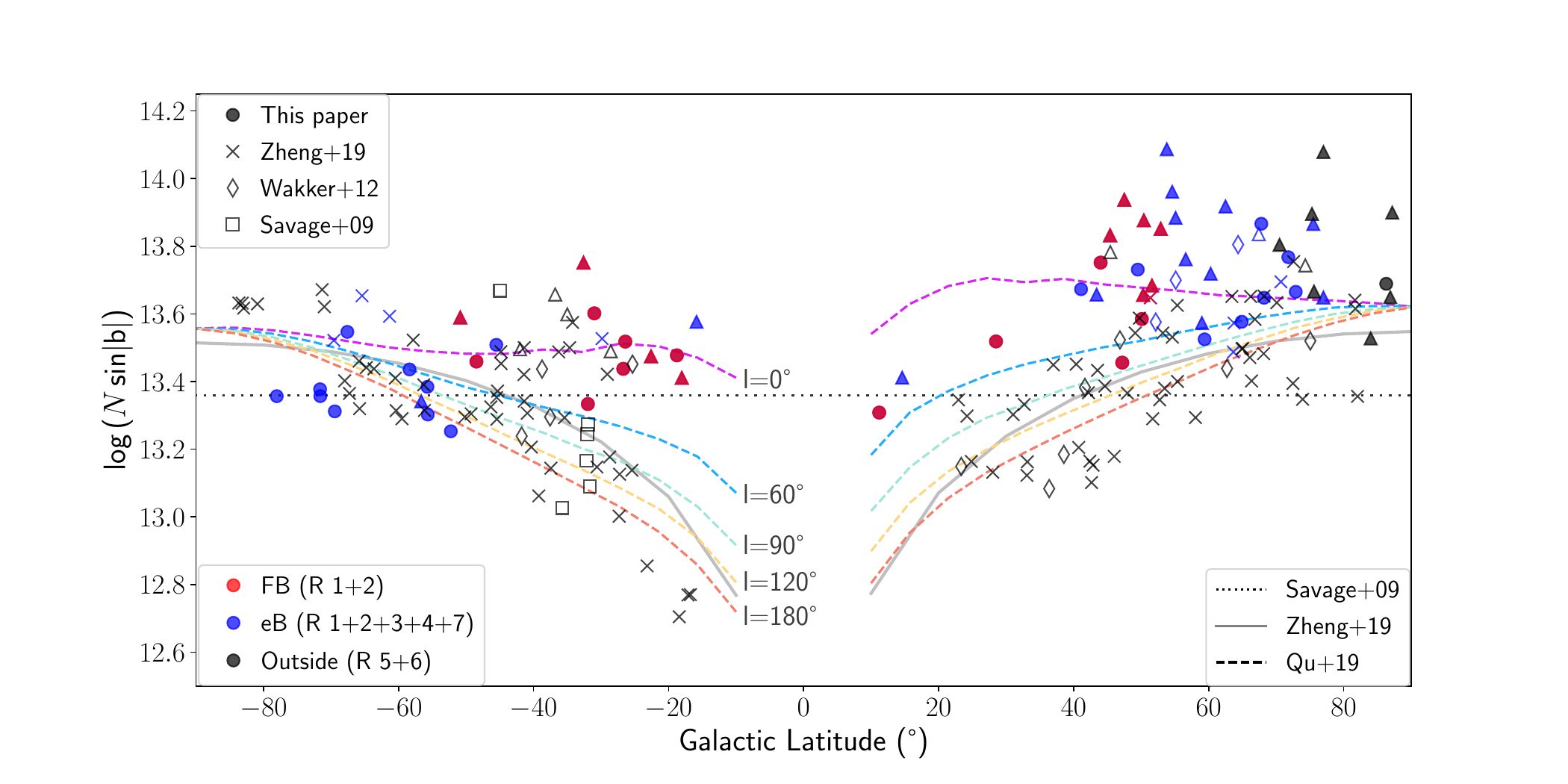}
    
\caption {\SiIV\ columns projected onto the z-axis, $\mathrm{log}(N_{a}\,\mathrm{sin}|b|)$, \frenchspacing{vs.} Galactic longitude (top) and Galactic latitude (bottom) for our sample (filled circles and triangles), the \citetalias{Zheng_2019} sample (crosses and empty triangles),  the \citetalias{Wakker_2012} sample (diamonds and empty triangles), and the \citetalias{Savage_2009} sample (squares and empty triangles). Red symbols depict sight lines inside the FBs, blue symbols depict sight lines inside the eBs, and black symbols depict sight lines outside of both bubbles. Triangles represent saturated values. In the top panel, the black, red, and blue  horizontal lines are the mean unsaturated values for lines outside of the bubbles (13.41), inside the FBs (13.55), and inside the eBs (13.58) for all samples excluding overlapping sight lines. In the bottom panel the black horizontal dotted line represents the flat-slab model predicted value \citepalias[13.36;][]{Savage_2009}, the grey solid line represents the predicted values from the \citetalias{Zheng_2019} model, and the dashed colored lines represent the \citet{Qu_2019} models for a variety of longitudes. 
\label{figure:lat_lon_plots}}
\end{figure*}


\subsection{Column-density dependence on Latitude and Longitude}

\citetalias{Savage_2009}, \citetalias{Wakker_2012}, and \citetalias{Zheng_2019} measured the LIV \SiIV\ absorption through the MW disk and CGM
with 31, 58, and 132 extragalactic sight lines covering the entire sky. \citetalias{Wakker_2012} and \citetalias{Zheng_2019} found that their LIV \SiIV\ column density measurements increase slightly by 0.1--0.26 dex in directions towards the GC regions, and \citetalias{Zheng_2019} suggest this enhancement could be due to the FBs.

To understand how the \SiIV\ measurements in our GC sample compare to \SiIV\ in the rest of the disk-halo interface and CGM, we combine the \citetalias{Savage_2009}, \citetalias{Wakker_2012}, and \citetalias{Zheng_2019} results with our own in Figure~\ref{figure:lat_lon_plots} where we plot $\mathrm{log}(N_{a}\,\mathrm{sin}|b|)$ \frenchspacing{vs.} Galactic latitude and longitude. $\mathrm{log}(N_{a}\,\mathrm{sin}|b|)$ measures the z-axis-projected column densities, where the z-axis perpendicular to the Galactic plane, and is designed to correct for disk projection effects. In a simple exponential disk model, $\mathrm{log}(N_{a}\,\mathrm{sin}|b|)$ is a constant
and thus becomes independent of Galactic latitude \citep{Savage_1990, Savage_2009, Zheng_2019}.

Of the 31, 58, and 132 sight lines from the \citetalias{Savage_2009}, \citetalias{Wakker_2012}, and \citetalias{Zheng_2019} samples, 3, 3, and 23 sight lines overlap with our sample, respectively. 
We compare the $\mathrm{log}(N_{a})$ measurements for the 23 overlapping \citetalias{Zheng_2019} sight lines and find they agree within the errors. We note that \citetalias{Wakker_2012} and \citetalias{Zheng_2019} use a stricter 1$\sigma$ criteria for defining matching doublet column density measurements, 
whereas our threshold is 3$\sigma$. We also note that \citetalias{Wakker_2012} integrates over a variable velocity range that is determined by the visual identification of the thick disk component.

In the top panel of Figure~\ref{figure:lat_lon_plots} we plot $\mathrm{log}(N_{a}\,\mathrm{sin}|b|)$ \frenchspacing{vs.} Galactic longitude for our sample and all three comparison samples. In the figure, the $\mathrm{log}(N_{a}\,\mathrm{sin}|b|)$ \SiIV\ values in the GC region stand out clearly against the rest of the Galactic halo, with GC sight lines having $\mathrm{log}(N_{a}\,\mathrm{sin}|b|)\gtrsim13.8$. 

\begin{figure*}[!ht]
    \centering
    \epsscale{1.15} 
    \plotone{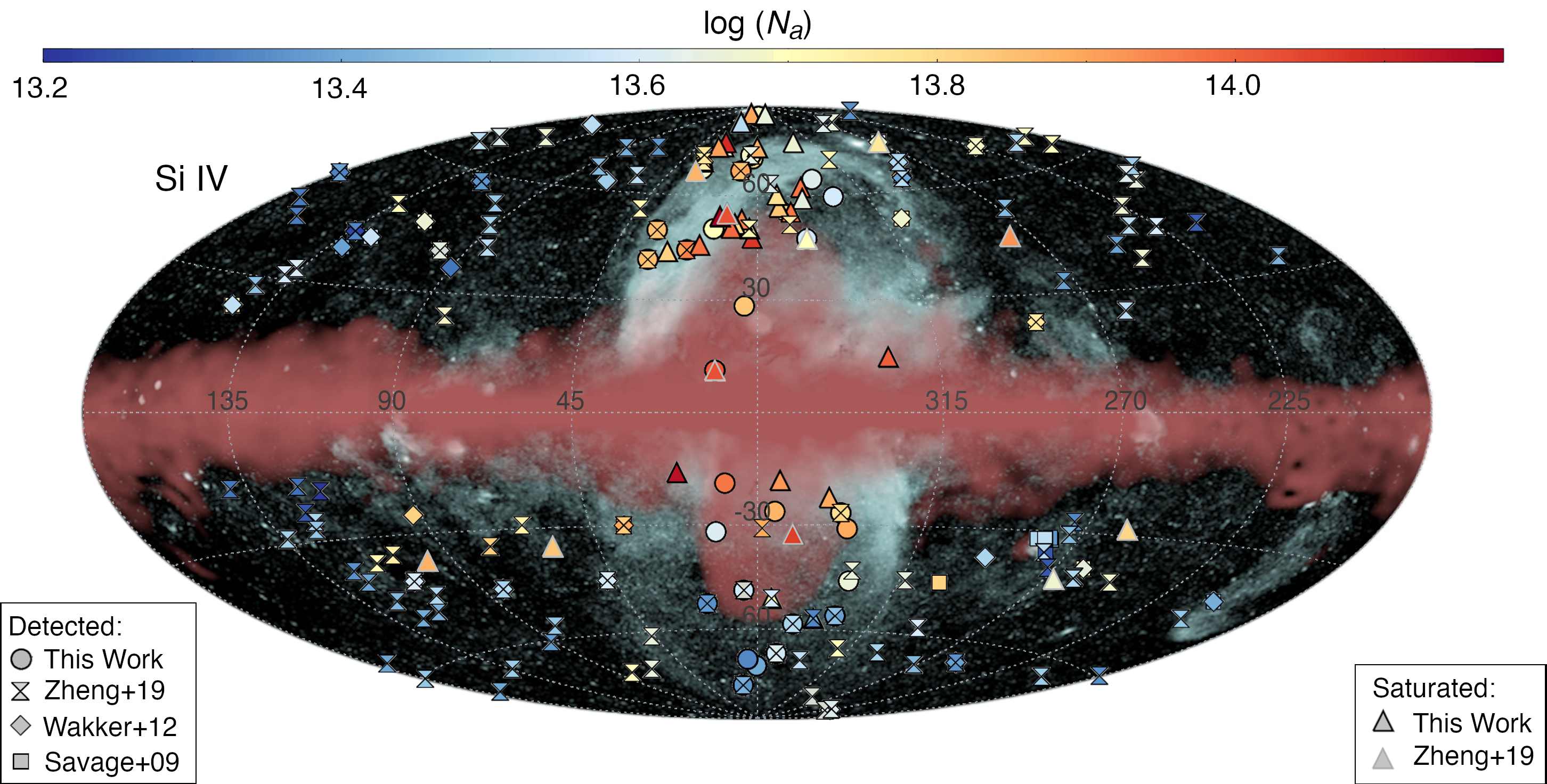}
\caption {Measured LIV log\,$N_{a}$(\SiIV) values ($N$ in cm$^{-2}$) from this paper,  \citetalias{Savage_2009}, \citetalias{Wakker_2012}, and \citetalias{Zheng_2019} projected onto the combined gamma-ray (red) and X-ray (cyan) map of the FBs and eBs adapted from \citet{Predehl_2020}. The LNPS association appears in this image as the strong X-ray emission in the top-left of the northern eB. 
\label{figure:predhel}}
\end{figure*}

To see this effect in Galactic latitude, we plot $\mathrm{log}(N_{a}\,\mathrm{sin}|b|)$ \frenchspacing{vs.} $b$. Figure~\ref{figure:lat_lon_plots} shows higher \SiIV\ column densities at positive latitudes, demonstrating an asymmetry between the northern and southern hemispheres. The northern enhancement in $\mathrm{log}(N_{a}\,\mathrm{sin}|b|)$ appears to occur within both the FB and eB latitude boundaries. 

In the bottom panel of Figure~\ref{figure:lat_lon_plots}, we plot the \SiIV\ column density predictions of three models of the disk-halo interface. First, we plot the \citetalias{Savage_2009} model which assumes that the disk column density decreases exponentially with a fixed scale height (plane-parallel slab model). Second, we plot the \citetalias{Zheng_2019} two-component model which adds a latitude-dependent CGM component to the plane-parallel slab model. Third we plot the \citet{Qu_2019} model which accounts for a two-dimensional radial distribution of the disk-halo interface with a latitude-dependent CGM. None of these models appear to account for the high columns in the northern sky. \citet{Qu_2019} and \citet{Qu_2020} also notice a north-south asymmetry in \SiIV\ columns and attempt to account for it by artificially increasing the northern sight lines absorption, increasing their model limits from $\mathrm{log}(N_{a}\,\mathrm{sin}|b|)\lesssim13.55$ to $\lesssim13.75$. Even with this increase in the modeled northern $\mathrm{log}(N_{a}\,\mathrm{sin}|b|)$ values, Figure~\ref{figure:lat_lon_plots} indicates that the GC northern enhancement in \SiIV\ measured columns cannot be explained from the shape of the disk-halo interface.

To understand exactly where these enhancements arise, we plot the measured $\mathrm{log}(N_{a})$ values of all samples on a map of the full sky X-ray and gamma-ray emission from \citet{Predehl_2020} in Figure~\ref{figure:predhel}. This figure shows both the FBs and eBs. In Figure~\ref{figure:predhel}, the eBs and FBs visually appear to have much higher \SiIV\ than the rest of the sky. 
We discuss the statistics of potential enhancements in \SiIV\ absorption  below in detail in  Section~\ref{section:Statistics}.

\begin{figure*}
    \centering
    \includegraphics[height=2.40cm]{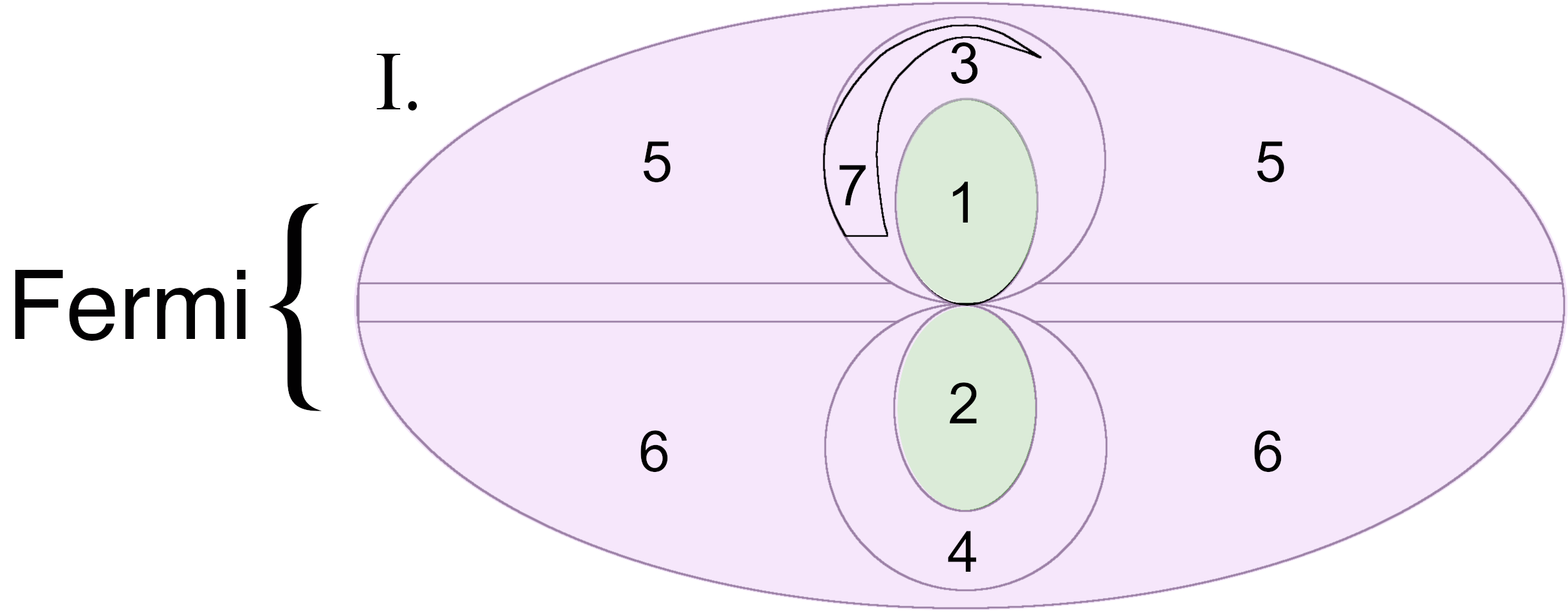}
    \includegraphics[height=2.40cm]{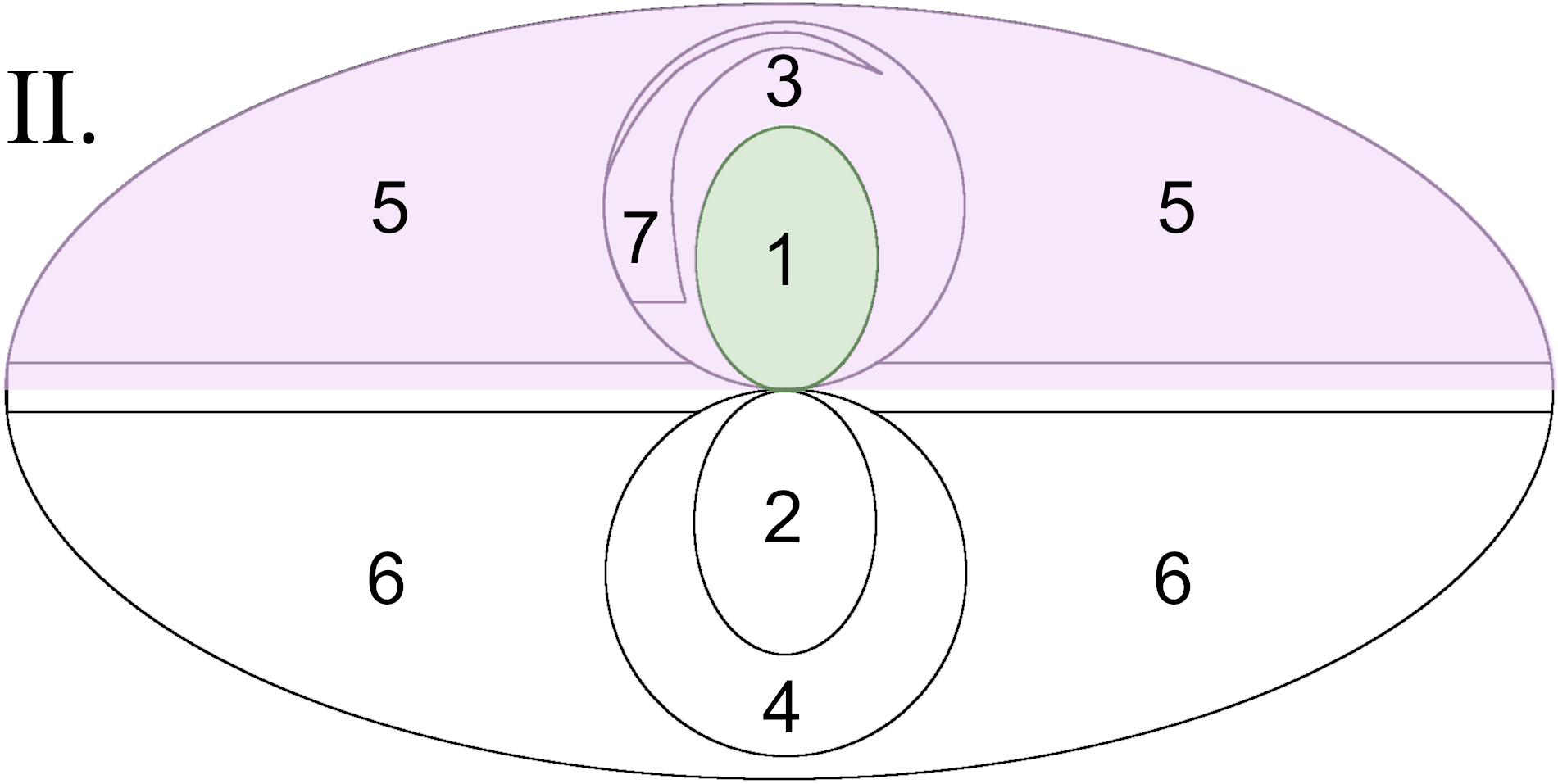}
    \includegraphics[height=2.40cm]{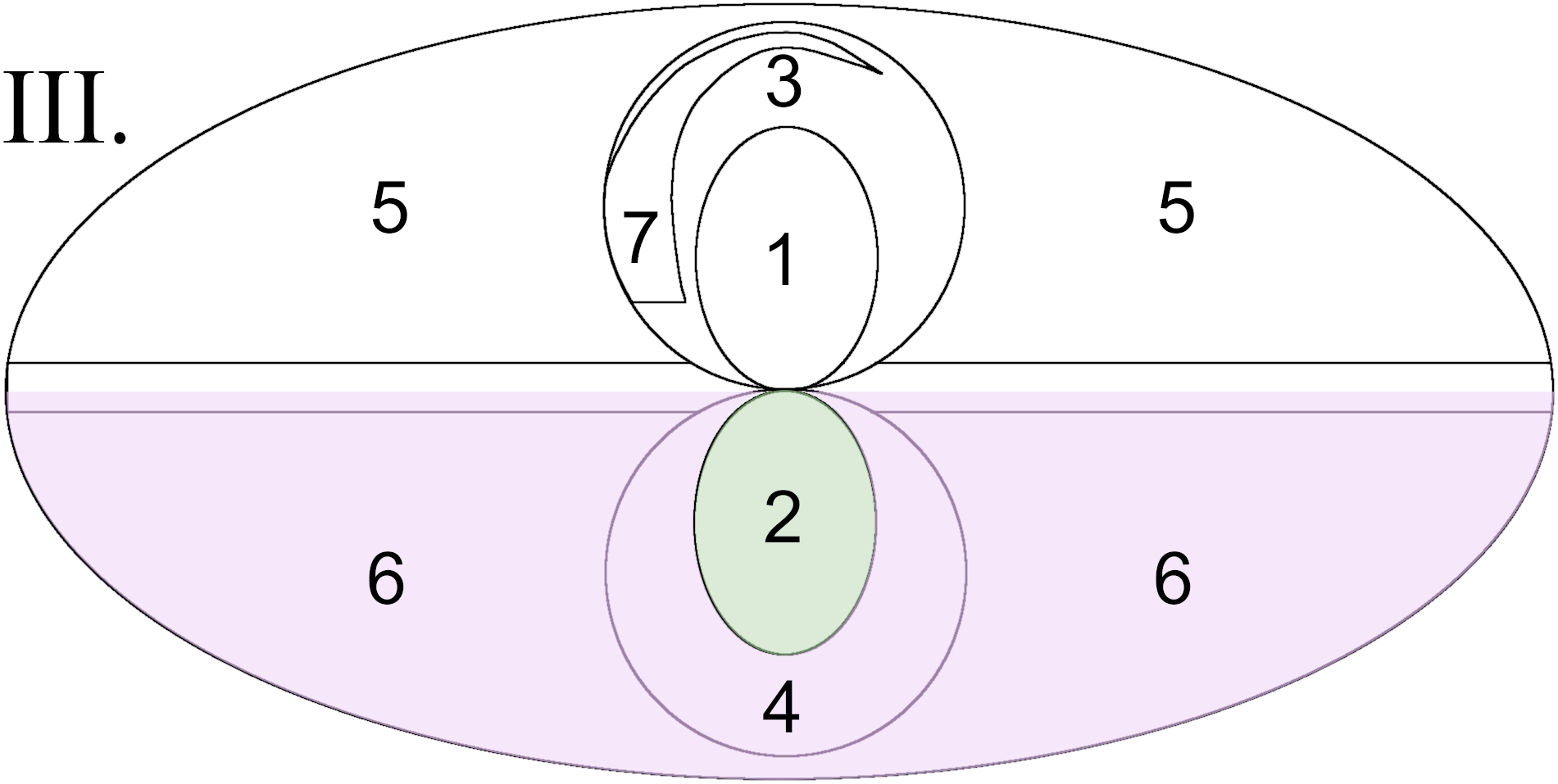}\\
    \includegraphics[height=2.40cm]{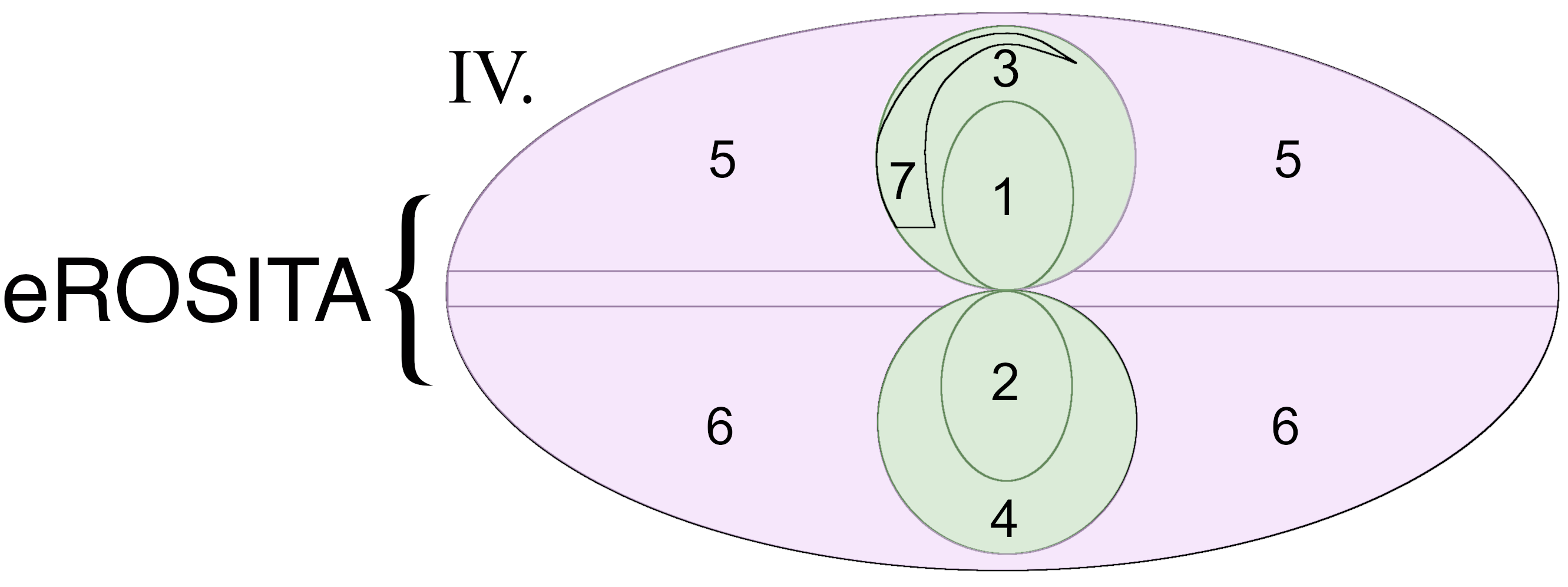}
    \includegraphics[height=2.40cm]{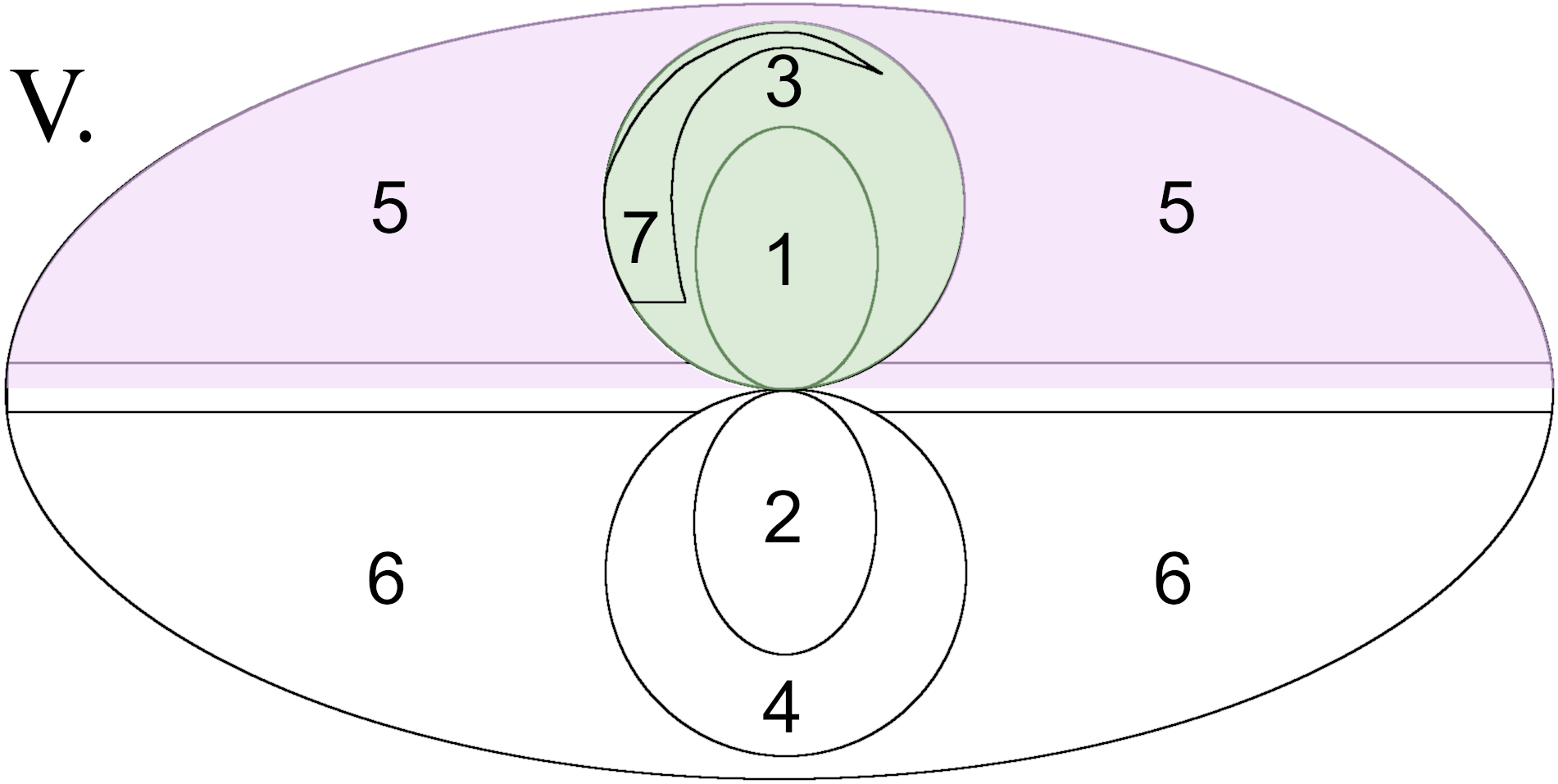}
    \includegraphics[height=2.40cm]{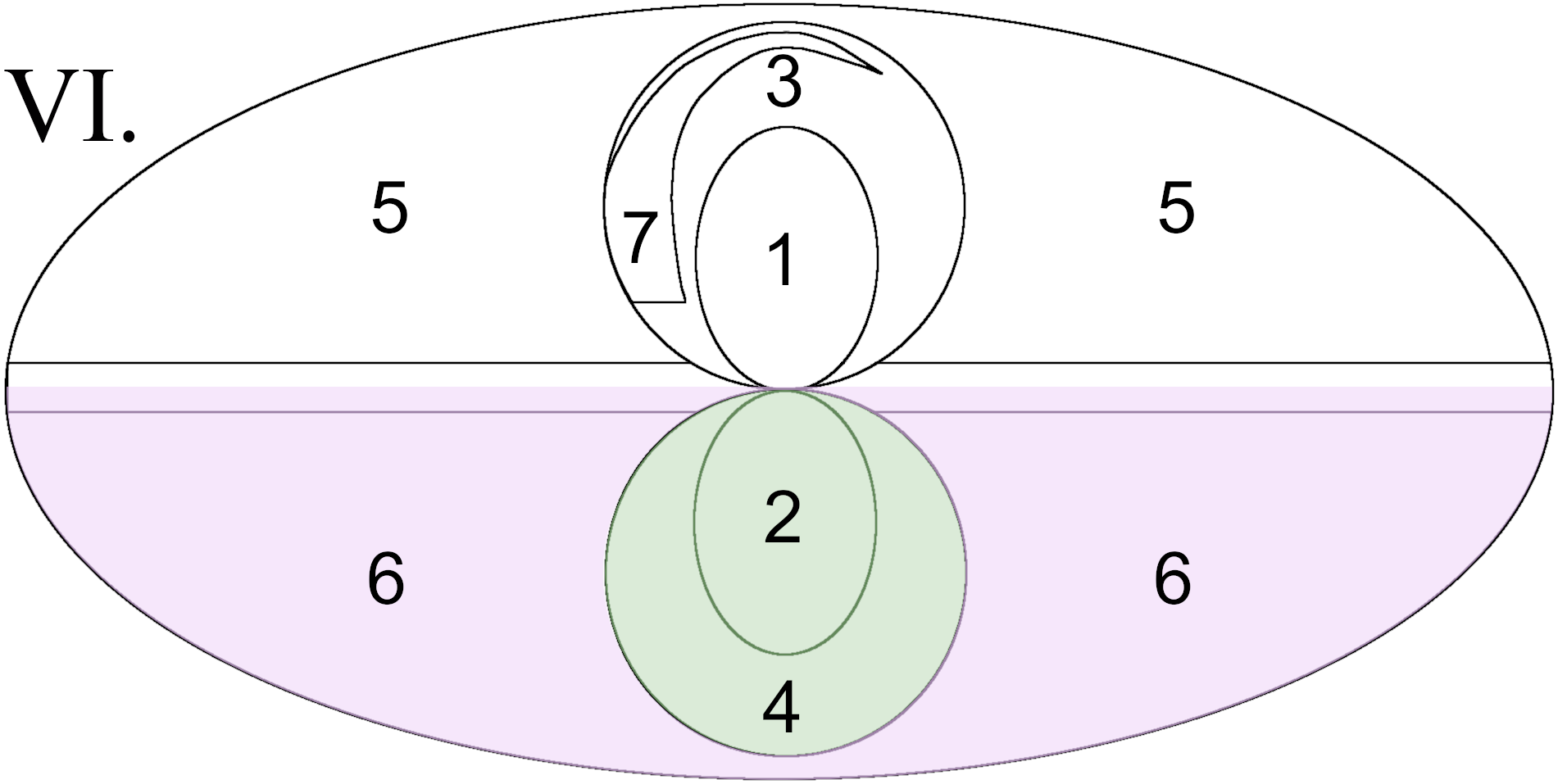}\\
    \includegraphics[height=2.40cm]{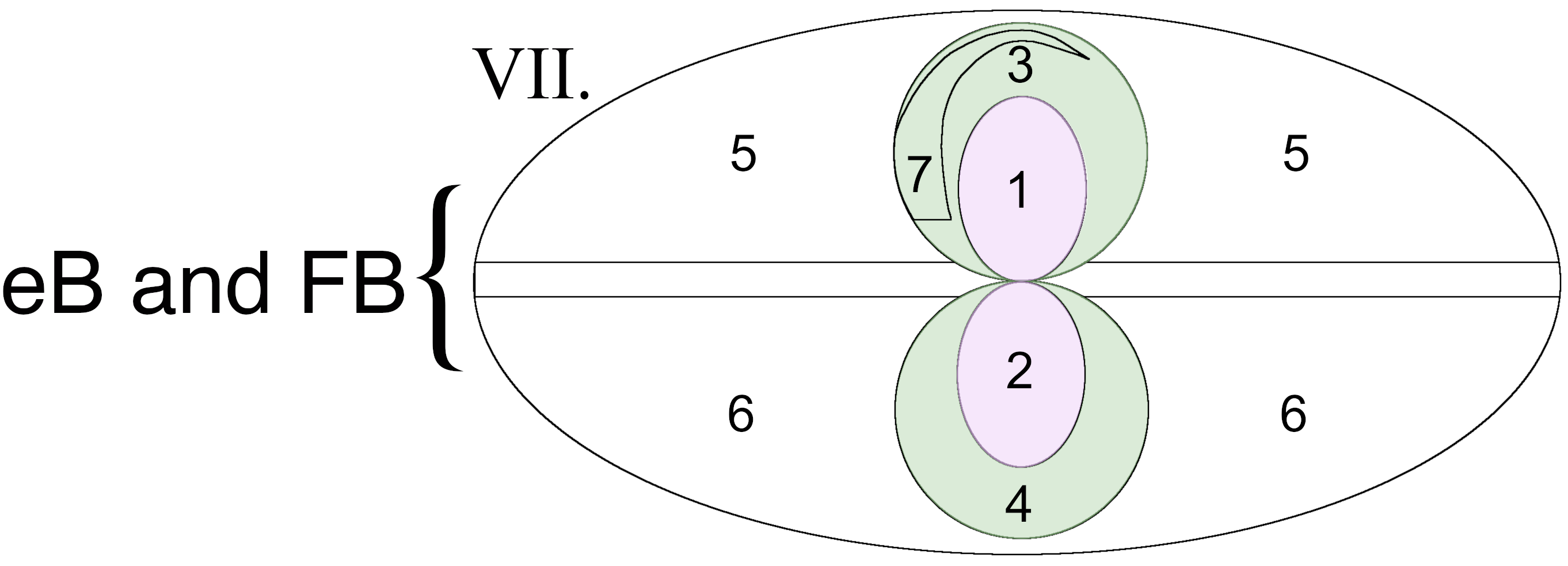}
    \includegraphics[height=2.40cm]{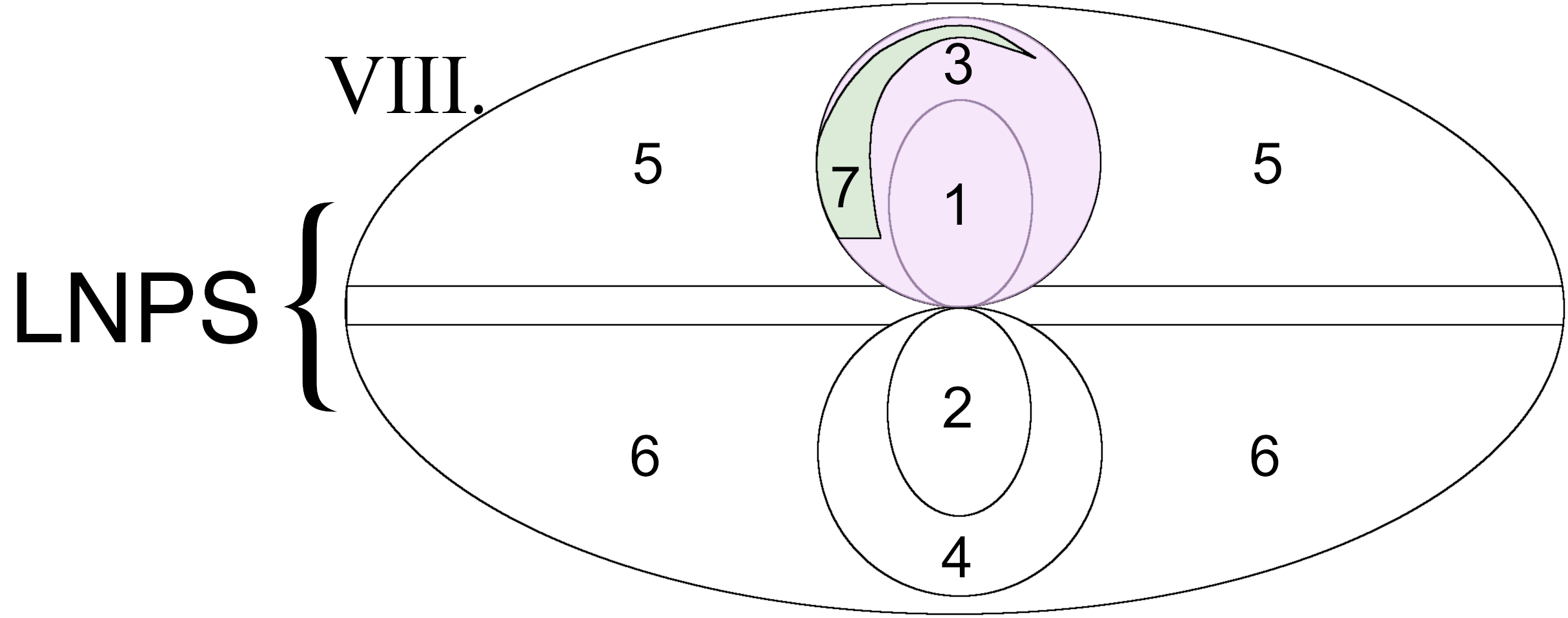}
    \caption{Maps of the sky regions explored in this study: FBs (1 and 2), eBs (3 and 4), sky regions exterior to the bubbles (5 and 6), and LNPS (7). The numbers and shading label the spatially-selected comparison groups referenced in Tables~\ref{table:N_measured_stats} and \ref{table:N_model_subtracted_stats}. 
\label{figure:regions}}
\end{figure*}

\begin{deluxetable*}{rclcccc|ccc}[!ht]
\setlength{\tabcolsep}{4pt}
\tablecaption{Statistical analysis of $N_{a}$(\SiIV) values combined from this work, \citetalias{Savage_2009}, \citetalias{Wakker_2012}, and \citetalias{Zheng_2019} \label{table:N_measured_stats}}
\tablehead{ & & \colhead{} & \colhead{} & \colhead{} & \multicolumn{2}{c}{\underline{~~~Basic Statistics~~~}} & \multicolumn{3}{c}{\underline{~Statistical Test~\tm{\footnotesize f}}}\\[-6pt]
 & \multicolumn{2}{c}{Spatially-Selected} &   \colhead{Count} & \colhead{Count} & \colhead{} & \colhead{Standard} & \colhead{} & \colhead{} \\[-6pt]
 & \multicolumn{2}{c}{Comparison Groups \tm{\footnotesize a}} &  \colhead{Unsaturated \tm{\footnotesize b}} & \colhead{Saturated \tm{\footnotesize c}} & \colhead{$\mathrm{log}(\overline{N_{a}})$ \tm{\footnotesize d}} &  \colhead{Error \tm{\footnotesize e}} &  \colhead{$\chi^{2}$} &  \colhead{$p$-value}  &  \colhead{Test\tm{\footnotesize g}}} 
\startdata  
\multirow{6}{*}{Fermi \Vast\{} & \multirow{2}{*}{\RN{1}.} & Inside FBs (R 1+2) & 12 & 10 & $13.93$  & 0.03 &  \multirow{2}{*}{\pmb{$20.1$}} & \multirow{2}{*}{\pmb{$7\times10^{-6}$}} & \multirow{2}{*}{P}\\
 &  & Outside FBs (R  3+4+5+6+7) &  137 & 28 & $13.71$ &  0.03 &  &   \\\cline{2-10}
 & \multirow{2}{*}{\RN{2}.} & Inside FB: North (R 1) & 6 & 6 & $13.94$ &  0.04 &  \multirow{2}{*}{\pmb{$8.5$}} & \multirow{2}{*}{\pmb{$0.004$}} & \multirow{2}{*}{P}\\ 
 &  & Outside FB: North (R 3+5+7) & 66 & 22 &  $13.79$ & 0.04 &   &   \\\cline{2-10}
 & \multirow{2}{*}{\RN{3}.} & Inside FB: South (R 2) &  6 & 4 & $13.89$ & 0.04 & \multirow{2}{*}{\pmb{14}} & \multirow{2}{*}{\pmb{$2\times10^{-4}$}}  & \multirow{2}{*}{L}\\
 &  & Outside FB: South (R 4+6) & 71  & 6 & $13.60$  & 0.03 &   &   \\\cline{2-10}
 \multirow{6}{*}{eROSITA \Vast\{} & \multirow{2}{*}{\RN{4}.} & Inside eBs (R 1+2+3+4+7) & 39 & 25 & $13.91$ & 0.03 & \multirow{2}{*}{\pmb{39}} & \multirow{2}{*}{\pmb{$4\times10^{-10}$}} & \multirow{2}{*}{L}\\ 
 &  & Outside eBs (R 5+6) & 110  & 13 & $13.62$ &  0.03 &  &  \\\cline{2-10}
 & \multirow{2}{*}{\RN{5}.} & Inside eB: North (R 1+3+7) & 19 & 19 & $13.96$ &  0.04 & \multirow{2}{*}{\pmb{32.9}} & \multirow{2}{*}{\pmb{$1\times10^{-8}$}}  & \multirow{2}{*}{P} \\
 &  & Outside eB: North (R 5) & 53  & 9 & $13.66$ &  0.04 &  &   \\\cline{2-10}
 & \multirow{2}{*}{\RN{6}.} & Inside eB: South (R 2+4) & 20 & 6 & $13.78$ & 0.06 & \multirow{2}{*}{\pmb{8.6}} & \multirow{2}{*}{\pmb{0.003}}  & \multirow{2}{*}{L} \\
 &  & Outside eB: South (R 6) & 57 & 4 & $13.56$ &  0.02 &   &  \\\cline{2-10}
 & \multirow{2}{*}{\RN{7}.} & Inside eBs (R 3+4+7) & 27 & 15 & $13.86$  & 0.05 & \multirow{2}{*}{\pmb{5.7}} & \multirow{2}{*}{\pmb{$0.02$}} & \multirow{2}{*}{P}\\
 &  & Inside FBs (R 1+2) & 12 & 10 & $13.93$  & 0.03 &   &  \\\cline{2-10}
 \multirow{2}{*}{LNPS \Big\{} & \multirow{2}{*}{\RN{8}.} & Inside LNPS (R 7) & 8 & 12 & 14.01 & 0.05 & \multirow{2}{*}{1.7} & \multirow{2}{*}{0.2} & \multirow{2}{*}{L}\\ 
 &   & Inside eB: North (R 1+3) & 11  & 7 & 13.87 &  0.05 &   & \\  %
\enddata
\tn{a}{Spatially-selected groups are compared two at a time with the comparison set bracketed by horizontal lines. The region (R) covered by these different group is denoted by a number shown in Figure~\ref{figure:regions}.}\vspace{-5pt}
\tn{b}{The number of unsaturated sight lines.}\vspace{-5pt}
\tn{c}{The number of saturated sight lines.}\vspace{-5pt}
\tn{d}{The log of the restricted mean apparent column density. The restricted mean is calculated using an upper limit set to the maximum column density in each spatially selected group.}\vspace{-5pt}
\tn{e}{The standard error on the log restricted mean.}\vspace{-5pt}
\tn{f}{Comparison sets with statistics in bold are likely drawn from separate populations.}\vspace{-5pt}
\tn{g}{Survival test used in the \texttt{survdiff} function: P = Peto-Peto test and L = Log-rank test.}
\end{deluxetable*}

\begin{deluxetable*}{rclcccc|ccc}[!ht]
\setlength{\tabcolsep}{4pt}
\tablecaption{Statistical analysis of disk-model-subtracted $N_{a}$(\SiIV) values combined from this work, \citetalias{Savage_2009}, \citetalias{Wakker_2012}, and \citetalias{Zheng_2019}  \label{table:N_model_subtracted_stats}}
\tablehead{ & &\colhead{} & \colhead{} & \colhead{} & \multicolumn{2}{c}{\underline{~~~Basic Statistics~~~}} & \multicolumn{3}{c}{\underline{~Statistical Test~\tm{\footnotesize f}}}\\[-6pt]
 &  \multicolumn{2}{c}{Spatially-Selected} & \colhead{Count} & \colhead{Count} & \colhead{} & \colhead{Standard} & \colhead{} & \colhead{} \\[-6pt]
 &  \multicolumn{2}{c}{Comparison Groups \tm{\footnotesize a}} & \colhead{Unsaturated \tm{\footnotesize b}} & \colhead{Saturated \tm{\footnotesize c}} & \colhead{$\mathrm{log}(\overline{N_{a, R}})$ \tm{\footnotesize d}} &  \colhead{Error \tm{\footnotesize e}} &  \colhead{$\chi^{2}$} &  \colhead{$p$-value} &  \colhead{Test\tm{\footnotesize g}}} 
\startdata  
\multirow{6}{*}{Fermi \Vast\{} & \multirow{2}{*}{\RN{1}.} & Inside FBs (R 1+2) & 12 & 10 & $13.18$  & 0.22 &  \multirow{2}{*}{$0.6$} & \multirow{2}{*}{$0.4$}  & \multirow{2}{*}{P}\\
 & & Outside FBs (R  3+4+5+6+7) &  137 & 28 & $13.10$ &  0.09 &  &   \\\cline{2-10}
 & \multirow{2}{*}{\RN{2}.} & Inside FB: North (R 1) & 6 & 6 & $13.16$ &  0.37 &  \multirow{2}{*}{$0.1$} & \multirow{2}{*}{$0.7$} & \multirow{2}{*}{P}\\ 
 & & Outside FB: North (R 3+5+7) & 66 & 22 &  $13.24$ & 0.10 &   &   \\\cline{2-10}
 &  \multirow{2}{*}{\RN{3}.} & Inside FB: South (R 2) &  6 & 4 & $13.18$ & 0.22 & \multirow{2}{*}{1.2} & \multirow{2}{*}{$0.3$}  & \multirow{2}{*}{L}\\
 & & Outside FB: South (R 4+6) & 71  & 6 & $12.76$  & 0.13 &   &   \\\cline{2-10}
 \multirow{6}{*}{eROSITA \Vast\{} & \multirow{2}{*}{\RN{4}.} & Inside eBs (R 1+2+3+4+7) & 39 & 25 & $13.42$ & 0.10 & \multirow{2}{*}{2.9} & \multirow{2}{*}{$0.09$} & \multirow{2}{*}{P} \\ 
 & & Outside eBs (R 5+6) & 110  & 13 & $12.91$ &  0.11 &  &  \\\cline{2-10}
 & \multirow{2}{*}{\RN{5}.} & Inside eB: North (R 1+3+7) & 19 & 19 & 13.56 &  0.11 & \multirow{2}{*}{\pmb{10.5}} & \multirow{2}{*}{\pmb{0.001}}  & \multirow{2}{*}{L}\\
 & & Outside eB: North (R 5) & 53  & 9 & 12.93 &  0.17 &  &   \\\cline{2-10}
 & \multirow{2}{*}{\RN{6}.} & Inside eB: South (R 2+4) & 20 & 6 & 12.79 & 0.29 & \multirow{2}{*}{0.0} & \multirow{2}{*}{1} & \multirow{2}{*}{L}\\
 & & Outside eB: South (R 6) & 57 & 4 & $12.83$ &  0.12 &   &  \\\cline{2-10}
 & \multirow{2}{*}{\RN{7}.} & Inside eBs (R 3+4+7) & 27 & 15 & $13.38$  & 0.12 & \multirow{2}{*}{0.2} & \multirow{2}{*}{$0.6$} & \multirow{2}{*}{L}\\
 & & Inside FBs (R 1+2) & 12 & 10 & $13.18$  & 0.22  &   &  \\\cline{2-10}
 \multirow{2}{*}{LNPS \Big\{} & \multirow{2}{*}{\RN{8}.} & Inside LNPS (R 7) & 8 & 12 & 13.67 & 0.10 & \multirow{2}{*}{\pmb{5.1}} & \multirow{2}{*}{\pmb{0.02}} & \multirow{2}{*}{P}\\ 
  &  & Inside eB: North (R 1+3) & 11  & 7 & 12.98 & 0.39 &   & \\  %
\enddata
\tn{a}{See Table~\ref{table:N_measured_stats} for table notes. Here the column densities have been disk-subtracted using the model of \citet{Qu_2019}.} 
\end{deluxetable*}\vspace{-48pt}

\subsection{Statistical Analysis of Spatially-Selected Regions}\label{section:Statistics}

To quantitatively assess the enhancement in \SiIV\ absorption towards the GC, we merge our sample with the \citetalias{Savage_2009}, \citetalias{Wakker_2012}, and \citetalias{Zheng_2019} samples and then divide it into eight spatially-selected comparison sets, as shown in Tables~\ref{table:N_measured_stats} and \ref{table:N_model_subtracted_stats} and Figure~\ref{figure:regions}. For the overlapping sight lines between our sample and comparison samples, we use our sample's measurements. We then choose which of the overlapping sight lines in the comparison samples to use based on the following priority: (1) \citetalias{Zheng_2019}, (2) \citetalias{Wakker_2012}, and (3) \citetalias{Savage_2009}. The only exception to this are the sight lines 3C273 and PKS0405-123, which appear in the \citetalias{Zheng_2019} as having unreliable \SiIV\ column measurements due to a $\lambda$1393 measurement that is abnormally stronger than the $\lambda$1402 measurement. For those sight lines, we use the \citetalias{Wakker_2012} value, who identified a reliable measurement when Lyman $\alpha$ interference and specific velocity ranges were taken into account. 

\begin{figure*}[!hb]
    \centering
    \epsscale{1.} 
    \plotone{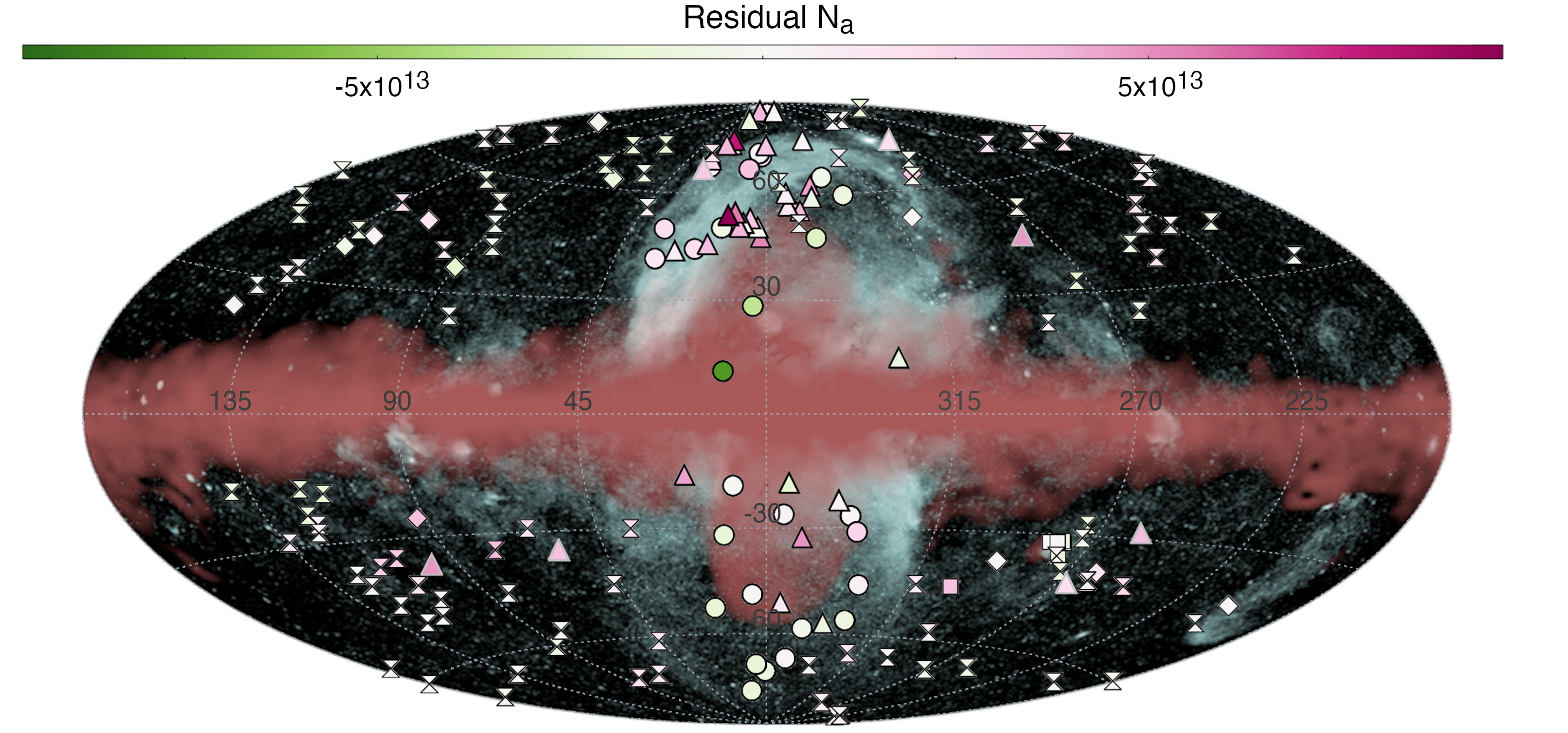}
    \plotone{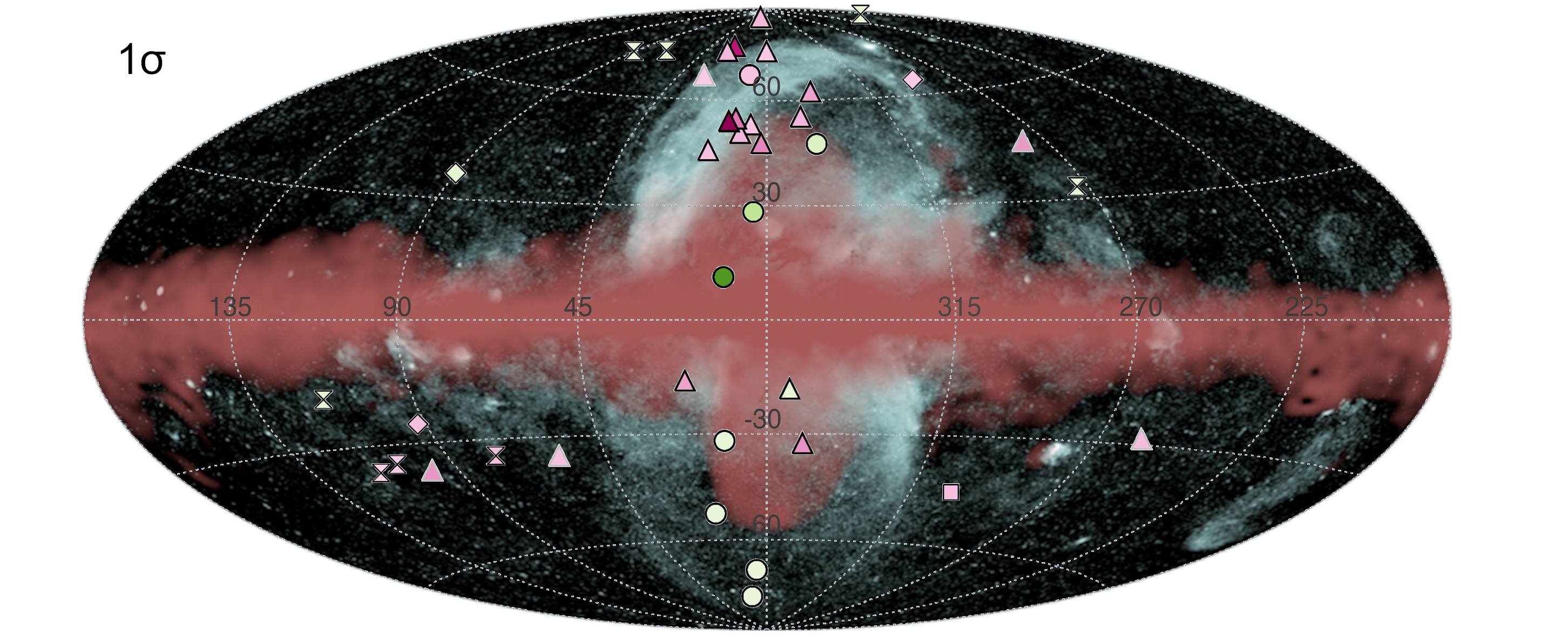}
    \plotone{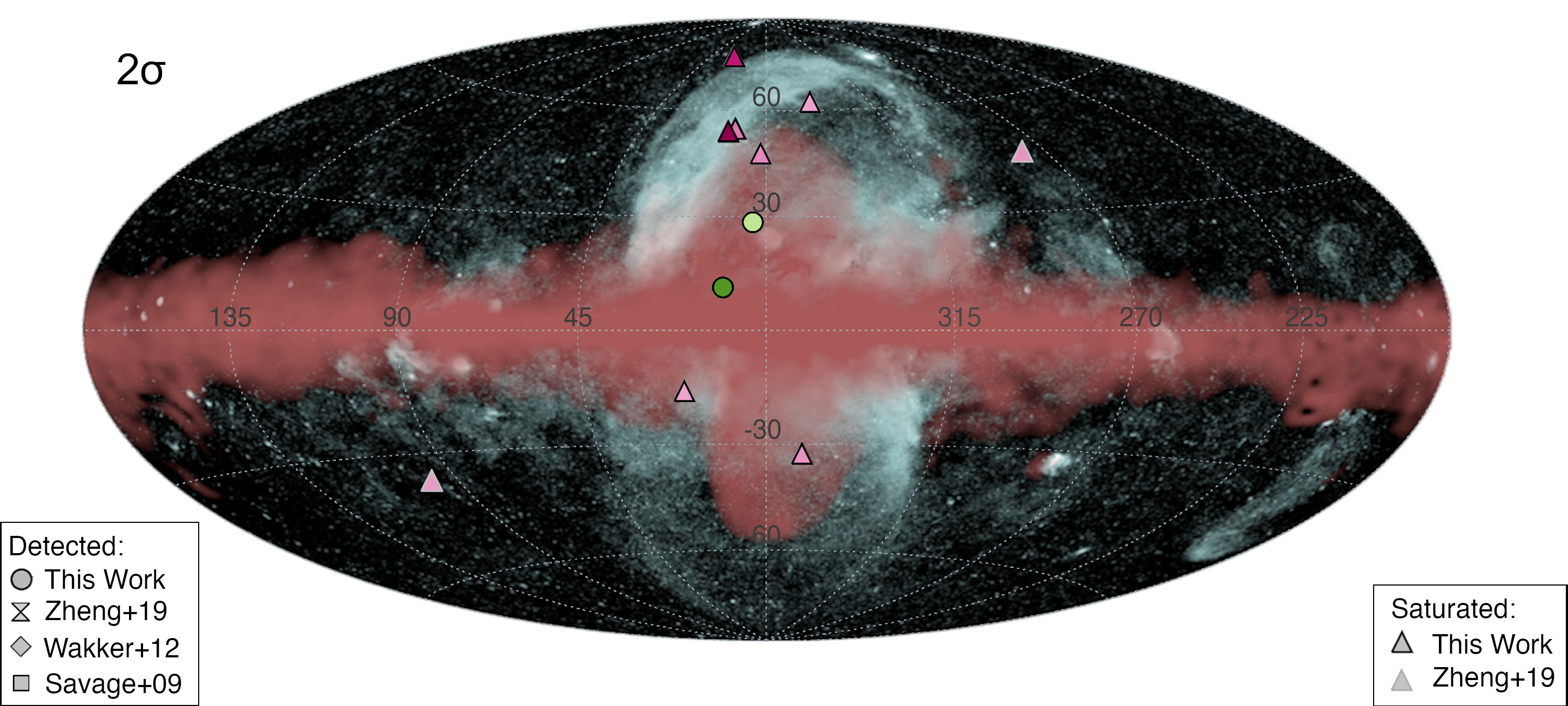}

\caption{Column-density residuals after the subtraction of the \citet{Qu_2019} disk model in the combined samples of this paper,  \citetalias{Savage_2009}, \citetalias{Wakker_2012}, and \citetalias{Zheng_2019}. Enhancements and deficits in densities are denoted in  green and pink, respectively. The top panel represents all sight lines, the middle and bottom panels represents all sightlines with residuals 1$\sigma$ and 2$\sigma$ away from the mean, respectively. \label{figure:residuals}}
\end{figure*}

For our statistical analysis, we analyze two sets of combined data: 
\begin{itemize}
\item The \SiIV\ column density measurements: analysis of the observed columns (see Figure~\ref{figure:predhel}) allows us to directly look for GC features that stand out compared to other measurements in a model-independent manner, 
without making assumptions about the shape of the disk. While statistically comparing the GC data to the entirety of the rest of disk-halo interface and CGM does not account for Galactic structure, our goal in this section is simply to measure the significance of the GC enhancement against the rest of the sky. This global approach is necessary because splitting the sky into different longitude or latitude bands for statistical tests provides samples that are too small to obtain statistically significant results. 

\item The disk-CGM model-subtracted \SiIV\ residuals: we analyze the residuals formed by subtracting the \citet{Qu_2019} two-dimensional disk-CGM \SiIV\ model from the \SiIV\ column measurements (see the top panel of Figure~\ref{figure:residuals}). The \citet{Qu_2019} model contains a radial- and height-dependent disk component and a constant CGM component, with \SiIV\ column densities determined by minimized $\chi^2$ models using AGN and stellar sight line measurements. For each of our sight lines we determine the model column at that sky position and subtract it from the observed column to obtain a residual. This model-dependent analysis  allows us to remove effects from the Galactic disk component on the column densities. These effects are particularly strong at low latitudes where sight lines pass through more of the ISM. 
\end{itemize}
Throughout the remainder of the paper, we will refer to these two datasets as the column density measurements and model-subtracted residuals, respectively.

Since a significant portion of the sight lines towards the GC are lower limits, we use survival analysis to account for censored data. 
Specifically, we use the  \texttt{survival} package function, \texttt{survfit} in the language \textsc{r}  to calculate each spatial group's restricted mean and its associated standard error \citep[see Tables~\ref{table:N_measured_stats} and \ref{table:N_model_subtracted_stats} for the column-density measurements and model-subtracted residual analysis, respectively;][]{R_2023, Therneau_2000, survival-package}. \texttt{survfit} uses the Kaplan-Meier method to create a survival curve for the data and then estimates the mean value using a range of acceptable columns \citep[a.k.a. the restricted mean, $\overline{N_{a, R}}$; ][]{Miller_1981}. For this analysis, we calculated the restricted mean by setting the maximum allowed column density in each spatially-selected group to the maximum column density of that specific spatial group. 

Additionally, we use \texttt{survdiff} function to determine if the two groups being compared are likely drawn from separate populations using either the log-rank or Peto-Peto test. The log-rank test gives equal weighting to all column density measurements and assumes proportional hazards \citep{Mantel_1966, Cox_1972}. The Peto-Peto test gives a larger weight to low column densities and is a more appropriate test for when the assumption proportional hazards is violated \citep{Gehan_1965, Peto_1972}. We decide which test to use by running the \texttt{cox.zph} function for each set of spatially selected comparison groups, a function that tests whether proportional hazards can be assumed \citep{Grambsch_1994, Therneau_2000}. In the results of both the log-rank or Peto-Peto tests, the $\chi^2$ value is an indication of how different the two survival curves are from what would be expected if they were drawn from the same population. The results also provide $p$-values which need to be $<0.05$ (95\% confidence) in order to reject the null hypothesis that the two spatially selected groups are drawn from the same population. For examples of the survival curves used in this analysis, see Appendix~\ref{section:survival_example}. 

All statistical tests were performed using the linear (not logarithmic) form of $N_{a}$. Survival analysis requires the values being evaluated to be positive, but the model-subtracted analysis results in both positive and negative residuals. As a solution, we found the minimum column residual in each pair of spatially compared groups and added the minimum residual to all residuals in both samples prior to running the statistical tests. This process ensures the survival analysis is conducted only on positive numbers and that the two compared groups are shifted by the same amount. After the tests, we then shift all restricted mean values by subtracting their respective minimum residual, accounting for our previous offset. The results of these tests are shown in Tables~\ref{table:N_measured_stats} and \ref{table:N_model_subtracted_stats}, and spatially selected groups that are likely drawn from separate populations are highlighted in bold.

\subsubsection{Sight Lines Inside vs. Outside the FBs}

Survival analysis tests indicate different populations for \textit{measured columns} in all spatially-compared groups associated with the FBs  ($p\le 0.004$). The sight lines passing though the FBs have a \SiIV\ $\mathrm{log}(\overline{N_{a}})$ value $0.22\pm0.04$ dex higher than the sight lines passing outside (Table~\ref{table:N_measured_stats}; Figure~\ref{figure:regions}: regions 1+2 vs. 3+4+5+6+7). This holds true in the northern Galactic hemisphere with a $0.15\pm0.06$ dex difference (Figure~\ref{figure:regions}: regions 1 vs. 3+5+7) and the southern Galactic hemisphere with a $0.39\pm0.05$ dex difference (Figure~\ref{figure:regions}: regions 2 vs. 4+6; see Table~\ref{table:N_measured_stats}).

The survival analysis tests for the \textit{model-subtracted residual} absorption do not show a difference in the populations for the Fermi Bubble (FB) spatially-selected comparison groups. Therefore, the location of the FB sight lines relative to the disk may be playing a large role in the excess of \SiIV\ columns in FB sight lines. 

\subsubsection{Sight Lines Passing Though vs. Outside the eBs}


The survival tests for the \textit{measured columns} indicate a difference in the population for all sight lines inside vs. outside the eBs (Figure~\ref{figure:regions}: regions 1+2+3+4+7 vs. 5+6; $p= 4\times10^{-10}$) and eB sight lines vs. non-eB sight lines in the southern Galactic hemisphere (regions 2+4 vs. 6; $p=0.003$).  On the other hand, their \textit{model-subtracted residual} absorption  do not show a difference in the populations of all and southern eB sight lines to non-eB sight lines. This could indicate that the disk is contributing to the differences in the populations. 

For sight lines passing through the \textit{northern} eB (Figure~\ref{figure:regions}: regions 1+3+7) both the measured $N_{a}$ and residual $N_{a, R}$ values have significantly different populations than northern directions outside the eB (Figure~\ref{figure:regions}: region 5), with $p$-values of $1\times10^{-8}$ and $1\times10^{-3}$ and differences of $0.30\pm0.06$ and $0.63\pm0.20$ dex, respectively. 

Between the measured and model-subtracted column density analysis, only the statistical tests for inside vs. outside the northern eROSITA Bubble show a difference in population in both Tables~\ref{table:N_measured_stats} and \ref{table:N_model_subtracted_stats}. This result indicates that there is strong absorption enhancement in the 
northern eB. 
The northern eB also has a model-subtracted $\mathrm{log}(\overline{N_{a, R}})$ value $0.77\pm0.39$ dex higher than the southern eB ($0.18\pm 0.07$ dex higher in the measured $\mathrm{log}(\overline{N_{a}})$; see Tables~\ref{table:N_measured_stats} and \ref{table:N_model_subtracted_stats}). We ran additional statistical tests for the eB north vs. south and found that the measured and model subtracted columns are likely drawn from different populations ($\chi^2=11.1$ and $6.1$, $p = 0.0009$ and $0.01$, respectively),
further highlighting the asymmetry in the northern and southern GC features seen in Figure~\ref{figure:lat_lon_plots}.

\subsubsection{Sight Lines Passing Though the FBs vs. eBs}

We continue to explore relationships of these different spatial regions by looking at sight lines that pass through the FBs (Figure~\ref{figure:regions}: regions 1+2) and sight lines passing through the eBs, but \textit{not} the FBs (Figure~\ref{figure:regions}: regions 3+4+7). Only the measured 
columns indicate that the populations are different ($p=0.02$). 
However, this difference may depend on the location of the sight lines with respect to the disk as we do not see a population difference in the model-subtracted survival test. 

\subsubsection{LNPS and Northern eB Sight Lines}\label{section:LNPS}

Strong \SiIV\ absorption is detected along the sight lines associated with the northern eB and LNPS association (Figure~\ref{figure:predhel}). To explore the specific role of the LNPS in the northern eB's LIV \SiIV\ enhancement, we separate out the sight lines passing through emission associated with LNPS. 

\begin{figure}[!hb]
    \centering
    \epsscale{1.1} 
    \plotone{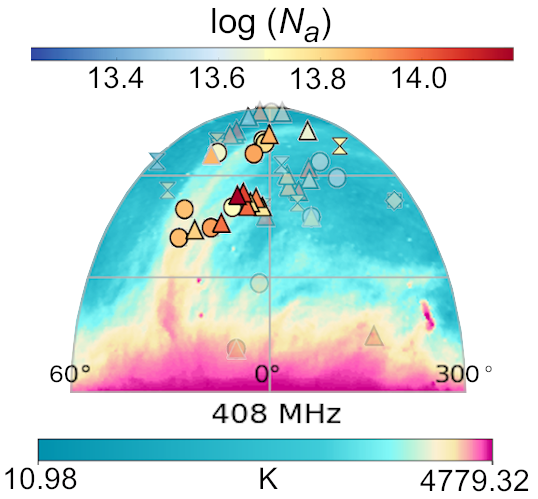}
\caption{Illustration of the location of the LNPS. 408 MHz continuum emission from \citet{Haslam_1982} and reprocessed by \citet{Remazeilles_2015} is shown in the color scale. Overplotted are the sight lines from this study,  \citetalias{Savage_2009}, \citetalias{Wakker_2012}, and \citetalias{Zheng_2019}, with the same labeling scheme as Figure~\ref{figure:predhel}. Sight lines that are considered to pass through the LNPS association for this study are opaque, while sight lines not included in the LNPS statistics for this study are transparent.  \label{figure:LNPS_regions}}
\end{figure}

In Figure~\ref{figure:LNPS_regions} we have plotted all of the sight lines selected as passing through the LNPS association in opaque symbols. These sight lines were selected based on the 408 MHz continuum emission \citep{Haslam_1982, Remazeilles_2015} and the Spektr-RG–eROSITA all-sky survey \citep{Predehl_2020}. We select the regions encompassed by LNPS based on definitions in the literature, such as \citet{Sofue_2015}, \citet{PlanckCollaboration_2016}, \citet{LaRocca_2020}, and references therein. As such, we select this sample based on LNPS's multi-wavelength appearance so that we include all sight lines that may pass through the LNPS association.  We chose not to include the faint southern structures that are at times associated with Loop I (Loop Is and the southern extension of Loop XII) because their connection with the northern LNPS association is not clear; it is possible that this structure is related to the southern eROSITA bubble or separate supernovae remnants \citep{Berkhuijsen_1971, Sofue_1979, PlanckCollaboration_2016, Predehl_2020, Panopoulou_2021, Lallement_2022}. 

We compare the LNPS to the rest of the  northern eB and do not find that the \textit{measured-column} populations are different from one another (Figure~\ref{figure:regions}: regions 7 vs. 1+3). However, the survival test for the \textit{model-subtracted} analysis does show a difference in the populations with a $p$-value of 0.02 and $\chi^2=5.1$. The LNPS has a model-subtracted $\mathrm{log}(\overline{N_{a, R}})$ enhancement of $0.69\pm0.40$ dex compared to the rest of the northern eB (see Table~\ref{table:N_model_subtracted_stats}). 

\subsection{\SiIV\ Enhancements towards the GC}\label{section:LNPS}

To better understand these enhancements in \SiIV, we have plotted sight lines in the model-subtracted analysis with residuals greater than 1$\sigma$ and 2$\sigma$ away from the mean in the middle and bottom panels of Figure~\ref{figure:residuals}, respectively. For this simple visual, we calculate the mean and standard deviation assuming that all values (including saturated values) are measurements, giving us a mean of $6.0\times10^{12}$ cm$^{-2}$ and a standard deviation of $1.9\times10^{13}$ cm$^{-2}$. From these plots we see several 1$\sigma$ enhancements (positive residuals) in sight lines near latitudes of $-30\degr$ and the strongest outliers detected near longitudes of 0$\degr$ with both enhancements and deficits (negative residuals) in sight lines. The 1$\sigma$ enhancements near the $-30\degr$ latitude line visually appear parallel to the disk and may reflect the inherent patchiness of the CGM and/or an underestimation of the disk contribution in the model at this latitude. The \SiIV\ enhancements and deficits in sight lines close to the $0\degr$ longitude line are so strong that they persist in the $2\sigma$ outliers plot on the bottom of panel in Figure~\ref{figure:residuals}. 
These 2$\sigma$ enhancements and deficits near $0\degr$ longitude could be an indication that the model does not reproduce accurate columns at low longitudes. 

To check how accurately the model reproduces the column densities at low longitudes, we use the \SiIV\ measurement from the sight line toward HD~167402, a star at the distance of $7.0\pm1.7$ kpc with $l, b = 2.26\degr,-6.39\degr$ that is towards the GC, but does not pass through GC features \citep[see ][]{Savage_2017}. This sight line has a measured LIV log\,$N$(\SiIV) of $13.65$ while the model predicts a log column of $13.90$, a difference of 0.25\,dex (\s$3.5\times 10^{13}$ cm$^{-2}$). This indicates that the model appears to be \textit{overestimating} the columns at low longitudes. If so, then the enhancements in the northern sight lines 
that are close to $0\degr$ longitude may be {\it even higher} than indicated in Figure~\ref{figure:residuals}. This would strengthen the conclusion from the statistical tests that there is a strong \SiIV\ enhancement towards the northern eB and LNPS.

If this excess absorption were associated with the Fermi or eBs alone, then we would expect these sight lines to be spread throughout both the northern and southern GC pointings. While a couple of enhanced southern GC sight lines are in the $2\sigma$ outlier plot, these two sight lines also fall close to the $-30\degr$ latitude line, which shows a general enhancement, 
possibly contributing to the enhancement in these two sight lines' residuals. 
Figure~\ref{figure:residuals}, combined with the statistical results in Tables~\ref{table:N_measured_stats} and \ref{table:N_model_subtracted_stats}, indicate that the northern eB has an enhancement of \SiIV. Since the LNPS occurs in the northern Galactic hemisphere but not the southern Galactic hemisphere, and most of the excess absorption is contained in the north, the excess absorption may be associated with the LNPS.  
Additionally, an enhancement towards the LNPS in FUV \CIV\ emission was found by \cite{Park_2007}, supporting an enhancement in high ions towards LNPS. The enhancement in the \SiIV\ LIV absorption suggests that the eB and  LNPS can be detected and characterized through excess \SiIV\ absorption. 

\section{Conclusions}\label{section:conclusion}

We have used archival HST/COS spectra to measure the apparent column densities of LIV \SiIV\ absorption along 61 AGN sight lines in the GC region (within $\pm$30$\degr$ in longitude of the GC). We have used these measurements to look for low-velocity signatures of high-ion absorption in three structures in the Galactic Center region: the FBs, eBs, and LNPS association. Using statistical tests that account for censored data,
we look for signatures of these features in the measured \SiIV\ column densities 
and in the residuals after the disk-CGM components are subtracted off, 
using the models of \citet{Qu_2019}. Our results are as follows:

\begin{enumerate}
\item 
{\bf Fermi Bubbles:} We find larger measured LIV \SiIV\ column densities 
in sight lines passing through the FBs than in sight lines passing outside, 
 with a mean enhancement of $0.22$ dex. However, when we subtract the disk-CGM components, a survival analysis test finds that the columns inside and outside the FBs are not drawn from different populations. This may indicate that the disk (foreground) \SiIV\ absorption is creating the differential between the two regions. 

\item 
{\bf eROSITA Bubbles:}  Survival tests between directions through the northern eB and the rest of the northern hemisphere reveal a significant difference in the distribution of LIV \SiIV\ column densities. This is true for both the measured and model-subtracted columns. The northern eB has \SiIV\ columns $0.30$ and $0.63$ dex higher than the rest of the northern hemisphere in measured and model-subtracted columns, respectively.


\item 
{\bf Northern enhancement:} The northern eB has a significantly higher mean \SiIV\ column than the southern eB for both the measured and model-subtracted analysis (0.18 and 0.77 dex higher, respectively), revealing a strong asymmetry between the northern and southern bubbles.
This can be seen in plots of the \SiIV\ residuals (observed -- model column densities), which show 
that \SiIV\ enhancements are clustered in the northern eB (Figure~\ref{figure:residuals}).

\item {\bf Loop I North Polar Spur (LNPS):}
If the \SiIV\ enhancement were primarily related to the eBs, then it would be expected in both hemispheres. The lack of a southern counterpart to this enhancement suggests that the LNPS (which lies in the north) may be the underlying source of the excess \SiIV\ LIV absorption. Statistical tests indicate an enhancement of 0.69 dex in the LNPS when compared to the rest of the northern eB. The LNPS enhancement is also supported by \ion{C}{4} emission measurements by \citet{Park_2007}.

\end{enumerate}


The results in this paper indicate a new method for detecting the northern eB and possibly the LNPS. The asymmetries detected in the \SiIV\ absorption reflect those seen in the X-ray, gamma-ray, and radio emission of the GC features. The source of the GC north-south brightness asymmetry and east-west asymmetry caused by the LNPS, is still a subject of debate. Some suggested sources include: variations in halo gas densities \citep{Sarkar_2018, Sofue_2019}, overlapping emission from a GC plasma bubble and a nearby supernova remnant \citep{Predehl_2020}, the outflow encountering the 3 kpc molecular ring in the GC \citep{Sofue_2021}, and CGM winds impacting the outflow \citep{Sofue_2019, Mou_2023}. The \SiIV\ measurements presented in this paper provide useful constraints for future models that attempt to understand the origins of the asymmetries in GC features.

\section{Acknowledgements}\label{section:acknowledgments}

We would like to thank the anonymous referee for their helpful comments. We would like to thank Zhijie Qu for providing us the disk-halo model from \citet{Qu_2019}. We gratefully acknowledge support from the NASA Astrophysics Data Analysis Program (ADAP) under grant 80NSSC20K0435, \textit{3D Structure of the ISM toward the Galactic Center} and from the STScI
Director's Discretionary Fund.
The HST COS data presented in this paper
were obtained from the Mikulski Archive for Space Telescopes (MAST) at the Space Telescope Science Institute. The specific observations analyzed for the sample in this paper can be accessed via MAST: \dataset[10.17909/jeph-7j87]{\doi{10.17909/jeph-7j87}}.

\clearpage

\bibliographystyle{aasjournal}

\bibliography{lvhi_Galactic_Center_4AGN}

\appendix
\restartappendixnumbering

\section{Sample LIV \SiIV\ Measurements}
We list basic information on the GC sample use for this study in Table~\ref{table:Soto}. We also list the log($N_a$) for the LIV \SiIV\ $\lambda\lambda$ 1393, 1402 , the adopted \SiIV\ log($N_a$), and the $N_{a,R}$ measurements.

\startlongtable
\setlength{\tabcolsep}{2pt}
\begin{deluxetable}{lccccccccc} 
\tabletypesize{\footnotesize}
\tablecaption{LIV \SiIV\ Measurements in COS AGN sight lines in the Galactic Center Region
\label{table:Soto}}
\tablehead{\colhead{Sight Line} & \colhead{$l$ (\degr)} & \colhead{$b$ (\degr)} & \colhead{Location} & \colhead{$\mathrm{log}\,N_{a,1393}$\tm{\footnotesize a}} & \colhead{$\mathrm{log}\,N_{a,1402}$\tm{\footnotesize b}} & \colhead{$\mathrm{log}\,N_{a,{\rm adop.}}$\tm{\footnotesize c}} & \colhead{$N_{a,R}$\tm{\footnotesize d}} & \colhead{PID\tm{\footnotesize e}} & \colhead{Ref.\tm{\footnotesize f}}}
\startdata  
RXJ1342.7+1844 & $0.24$ & $75.52$ & eB, LNPS & $\ge13.70\pm0.03$ & $\ge13.88\pm0.04$ & $\ge13.88$  & $\ge3.1\times10^{13}$ & 12248 & 2  \\
HE2332-3556 & 0.59 & $-$71.59 & eB & $13.31\pm0.06$ & $13.44\pm0.09$ & $13.38\pm0.12$ & $-1.2\times10^{13}$ &  13444 & 3 \\
RBS2023 & 0.61 & $-$71.62 & eB & $13.40\pm0.04$ &  ND\tm{\footnotesize g} & $13.40\pm0.04$   & $-1.1\times10^{13}$  & 13444 & 3\\
SDSSJ151237.15+012846.0 & 1.80 & 47.50 & FB, eB & $\ge13.98\pm0.03$ & $\ge14.07\pm0.05$ & $\ge14.07$ & $\ge5.3\times10^{13}$  & 12603 & 2 \\
MRK1392 & 2.80 & 50.30 & FB, eB, LNPS & $\ge13.68\pm0.01$ & $13.77\pm0.02$ & $\ge13.77$ & $\ge-2.2\times10^{12}$ &  13448 & 2\\
SDSSJ135712.60+170444.0 & 2.90 & 71.80 & eB, LNPS & $\le13.80\pm0.01$\tm{\footnotesize h} & $13.79\pm0.02$ & $13.79\pm0.02$  &  $1.6\times10^{13}$ &  12248 & 2 \\
1H1613-097 & 3.50 & 28.50 & FB, eB & $\ge13.76\pm0.02$ & $13.84\pm0.03$ & $13.84\pm0.03$  &  $-3.4\times10^{13}$ & 13448 & 2 \\
PG1352+183 & 4.40 & 72.90 & eB, LNPS & $13.68\pm0.01$ & $13.69\pm0.02$ & $13.69\pm0.02$  & $2.4\times10^{12}$ &  13448 & 2 \\
RBS1768 & 4.51 & $-$48.46 & FB, eB & $13.59\pm0.02$ & $13.58\pm0.03$ & $13.59\pm0.04$  & $-1.5\times10^{12}$ &  12936 & 3\\
CTS487 & 5.54 & $-$69.44 & eB & $13.36\pm0.03$ & $13.32\pm0.07$ & $13.34\pm0.07$  & $-1.4\times10^{13}$ & 13448 & 3\\
RBS1454 & 5.60 & 52.90 & FB, eB, LNPS & ...\tm{\footnotesize h} & $\ge13.95\pm0.03$ & $\ge13.95$  & $\ge3.1\times10^{13}$ & 12603 & 2 \\
SDSSJ150928.30+070235.0 & 7.80 & 51.60 & FB, eB, LNPS & $\ge13.85\pm0.02$\tm{\footnotesize i} & $\ge13.79\pm0.05$ &  $\ge13.79$ & $\ge2.4\times10^{12}$  & 12603 & 2  \\
UVQSJ191928.05-295808.0 & 8.18 & $-$18.77 & FB, eB & ...\tm{\footnotesize h} & $13.97\pm0.01$ &  $13.97\pm0.01$ & $4.3\times10^{11}$ &  15339 & 4\\
SDSSJ141542.90+163413.7 & 8.80 & 67.80 & eB, LNPS & $\ge13.85\pm0.01$ & $13.90\pm0.02$ & $13.90\pm0.02$  & $3.2\times10^{13}$ &  12486 & 2 \\
SDSSJ151507.43+065708.3 & 9.00 & 50.40 & FB, eB, LNPS & $\ge13.93\pm0.03$ & $\ge13.99\pm0.04$ & $\ge13.99$  & $\ge3.8\times10^{13}$ & 12603 & 2  \\
PDS456 & 10.40 & 11.20 & FB, eB &  $\ge14.05\pm0.01$ & $14.02\pm0.02$ & $14.02\pm0.02$  &  $-7.6\times10^{13}$ & 13448 & 1,2 \\
MRK841 & 11.20 & 54.60 & eB, LNPS & $\ge13.91\pm0.01$ & $14.05\pm0.01$ & $\ge14.05$  & $\ge5.7\times10^{13}$ &  13448 & 2\\
ESO462-G09 & 11.33 & $-$31.95 & FB, eB & $\ge13.62\pm0.03$ & $13.61\pm0.07$ &  $13.61\pm0.07$  & $-1.6\times10^{13}$  & 13448 & 3\\
RBS2070 & 12.84 & $-$78.04 & eB & $13.33\pm0.04$ & $13.40\pm0.07$ & $13.37\pm0.08$  & $-1.3\times10^{13}$ & 12864 & 3\\
SDSSJ150952.19+111047.0 & 13.60 & 53.80 & eB, LNPS & $\ge13.79\pm0.02$ & $\ge14.18\pm0.02$ & $\ge14.18$  & $\ge9.6\times10^{13}$ & 12614 & 2  \\
PG1522+101 & 14.90 & 50.10 & FB, eB, LNPS & ...\tm{\footnotesize h} & $13.70\pm0.02$ & $13.70\pm0.02$ & $-8.0\times10^{12}$ & 11741 & 2\\
PKS2155-304 & 17.73 & $-$52.25 & eB & $13.36\pm0.02$ & $13.35\pm0.03$ & $13.36\pm0.04$  & $-1.4\times10^{13}$ & 8125/12038 & 3 \\
SDSSJ154553.48+093620.5  & 18.30 & 45.40 & FB, eB, LNPS & $\ge13.98\pm0.02$ & ...\tm{\footnotesize h} &  $\ge13.98$ & $\ge3.3\times10^{13}$ & 12248 & 2 \\
UVQSJ192636.95-182553.0 & 20.00 & $-$15.86 & eB & $\ge14.14\pm0.01$ & $\ge14.14\pm0.01$ & $\ge14.14$ & $\ge4.5\times10^{13}$ & 15339 & 4\\
RXJ1303.7+2633 & 21.80 & 87.20 & - & $\ge13.90\pm0.03$ & ...\tm{\footnotesize h} & $\ge13.90$ & $\ge3.7\times10^{13}$ &  13382 & 2\\
PG1553+113 & 21.90 & 44.00 & FB, eB, LNPS & ...\tm{\footnotesize h} & $13.91\pm0.01$ & $13.91\pm0.01$  & $1.9\times10^{13}$ & 11520/12025 & 2\\
SDSSJ141038.39+230447.1 & 24.60 & 71.60 & eB & ...\tm{\footnotesize h} & ...\tm{\footnotesize h} & ...  & ... & 12958 & 2  \\
SDSSJ135424.90+243006.3 & 25.90 & 75.60 & - & $\ge13.68\pm0.04$ & ...\tm{\footnotesize h} & $\ge13.68$ & $\ge3.5\times10^{12}$ & 12603 & 2 \\
SDSSJ134822.31+245650.1 & 26.40 & 77.00 & - & $\ge14.00\pm0.03$ & $\ge14.09\pm0.05$ & $\ge14.09$  & $\ge7.9\times10^{13}$ & 12603 & 2 \\
RXJ1605.3+1448 & 27.80 & 43.40 & eB, LNPS & $\ge13.82\pm0.01$ & ...\tm{\footnotesize h} & $\ge13.82$  & $\ge6.8\times10^{12}$ &  12614 & 2 \\
SDSSJ131802.01+262830.3 & 28.20 & 84.00 & - & ...\tm{\footnotesize h} & $\ge13.53\pm0.15$ & $\ge13.53$ & $\ge-9.0\times10^{12}$ & 12603 & 2 \\
RXJ1356.4+2515 & 29.30 & 75.30 & - & $\ge13.83\pm0.02$ & $\ge13.91\pm0.02$ & $\ge13.91$ & $\ge3.7\times10^{13}$ & 12248 & 2 \\
PG1424+240 & 29.50 & 68.20 & eB, LNPS & ...\tm{\footnotesize h} & $13.68\pm0.03$ & $13.68\pm0.03$  & $2.3\times10^{12}$ & 12612 & 2  \\
NGC5548 & 32.00 & 70.50 & - & $\ge13.77\pm0.01$ & $13.83\pm0.01$  & $\ge13.83$ &  $\ge2.3\times10^{13}$ & 7572 & 2 \\
MRK877 & 32.90 & 41.10 & eB, LNPS & $13.84\pm0.01$ & $13.87\pm0.02$ & $13.86\pm0.02$ & $1.2\times10^{13}$ & 12569 & 2 \\
3C323.1 & 33.90 & 49.50 & eB, LNPS & $\ge13.83\pm0.01$ & $13.85\pm0.03$ & $13.85\pm0.03$ & $1.9\times10^{13}$ & 13398 & 2\\
QSO1503-4152 & 327.70 & 14.60 & eB & $\ge13.91\pm0.01$ & $\ge14.01\pm0.03$ & $\ge14.01$ &  $\ge-1.1\times10^{13}$ & 11659 & 2 \\
HE1340-0038 & 328.80 & 59.40 & eB & $13.46\pm0.05$ & $13.69\pm0.06$ & $13.59\pm0.08$ & $-9.3\times10^{12}$ & 11598/13033 & 2\\
SDSSJ131545.21+152556.3 & 329.90 & 77.00 & eB, LNPS & ...\tm{\footnotesize h} & $\ge13.66\pm0.08$ & $\ge13.66$ & $\ge1.7\times10^{12}$ & 12603 & 2\\
HE2259-5524 & 330.64 & $-$55.72 & eB & $13.51\pm0.03$\tm{\footnotesize i} & $13.42\pm0.07$\tm{\footnotesize i} & $13.47\pm0.07$  & $-5.8\times10^{12}$ & 13444 & 3 \\
HE2258$-$5524 & 330.72 & $-$55.67 & eB & $13.36\pm0.03$\tm{\footnotesize i} & $13.41\pm0.05$\tm{\footnotesize i} & $13.39\pm0.06$   & $-1.1\times10^{13}$ & 13444 & 3 \\
IRASF21325-6237 & 331.14 & $-$45.52 & eB & $13.64\pm0.01$ & $13.67\pm0.02$ & $13.66\pm0.02$ &  $7.1\times10^{12}$ & 12936  & 3\\
RXJ1342.1+0505 & 333.90 & 64.90 & eB & $\ge13.58\pm0.04$ & $13.62\pm0.07$ & $13.62\pm0.07$ & $-5.4\times10^{12}$ & 12248 & 2\\
SDSSJ125846.66+242739.1 & 335.10 & 86.90 & - & $\ge13.65\pm0.03$ & ...\tm{\footnotesize h} & $\ge13.65$  &  $\ge2.2\times10^{12}$ & 13382 & 2 \\
ESO141-G55 & 338.18 & $-$26.71 & FB, eB & $13.79\pm0.01$ & $13.78\pm0.01$ & $13.79\pm0.01$ & $1.5\times10^{12}$ & 12936 & 3\\
RBS1897 & 338.51 & $-$56.63 & eB & $\ge13.30\pm0.02$ & $13.42\pm0.03$ & $\ge13.42$  & $\ge-9.4\times10^{12}$  & 11686 & 3\\
SDSSJ135726.27+043541.4 & 340.80 & 62.50 & eB & ...\tm{\footnotesize h} & $\ge13.97\pm0.02$ & $\ge13.97$ &  $\ge4.4\times10^{13}$ & 12264 & 2 \\
UVQSJ185649.37-544229.9 & 341.66 & $-$22.60 & FB, eB & $\ge13.88\pm0.02$ & $\ge13.89\pm0.04$ & $\ge13.89$ & $\ge5.5\times10^{12}$  & 15339 & 4\\
SDSSJ140655.66+015712.8 & 341.80 & 59.00 & eB & ...\tm{\footnotesize h} & $\ge13.64\pm0.10$ & $\ge13.64$  & $\ge-7.2\times10^{12}$ & 12603 & 2  \\
PG1435-067 & 344.00 & 47.20 & FB, eB & $13.57\pm0.02$ & $13.61\pm0.04$ & $13.59\pm0.05$  & $-2.3\times10^{13}$ & 12569/13448 & 2\\
RBS1892 & 345.90 & $-$58.37 & eB & $13.48\pm0.02$\tm{\footnotesize i} & $13.53\pm0.04$\tm{\footnotesize i} & $13.51\pm0.05$   & $-4.0\times10^{12}$ & 12604 & 3\\
SDSSJ142614.79+004159.4 & 347.60 & 55.10 & eB & $\ge13.96\pm0.03$ & $\ge13.97\pm0.04$ & $\ge13.97$ & $\ge3.9\times10^{13}$  & 13473 & 2 \\
RBS2000 & 350.20 & $-$67.58 & eB & $13.55\pm0.02$ & $13.61\pm0.04$ & $13.58\pm0.05$ & $2.4\times10^{12}$ & 13448 & 3\\
PKS2005-489 & 350.37 & $-$32.60 & FB, eB & $\ge13.97\pm0.01$ & $14.02\pm0.01$ & $\ge14.02$  & $\ge4.9\times10^{13}$ & 11520 & 3\\
RXJ1429.6+0321 & 351.80 & 56.60 & eB & $\ge13.77\pm0.04$ & $\ge13.84\pm0.06$ & $\ge13.84$ & $\ge1.5\times10^{13}$  &  12603 & 2\\
SDSSJ141949.39+060654.0 & 351.90 & 60.30 & eB & $\ge13.78\pm0.03$ & ...\tm{\footnotesize h} & $\ge13.78$ & $\ge8.7\times10^{12}$  & 13473 & 2\\
UVQSJ185302.65-415839.6 & 354.36 & $-$18.04 & FB, eB & $\ge14.03\pm0.01$ & $13.92\pm0.02$ & $\ge13.92$ & $\ge-1.7\times10^{13}$  & 15339 & 4\\
RBS1795 & 355.18 & $-$50.86 & FB, eB & $\ge13.54\pm0.01$ & $13.70\pm0.02$ & $\ge13.70$  &  $\ge1.1\times10^{13}$ & 11541 & 3\\
UVQSJ193819.59-432646.3 & 355.47 & $-$26.41 & FB, eB & $\ge13.86\pm0.03$\tm{\footnotesize h} & $13.87\pm0.02$\tm{\footnotesize h} & $13.87\pm0.02$ & $3.8\times10^{12}$ & 15339 & 4\\
HS1302+2510 & 357.40 & 86.30 & - & $\ge13.74\pm0.02$ & $13.69\pm0.06$ & $13.69\pm0.06$ &  $6.4\times10^{12}$ & 13382 & 2\\
RBS1666 & 335.73 & $-$31.00 & FB, eB & $\ge13.89\pm0.01$ & $13.89\pm0.02$ & $13.89\pm0.02$ & $2.5\times10^{13}$ & 13448	& 3\\ 
\enddata 
\tn{a}{The log of the apparent column density ($N_a$ in cm$^{-2}$) measured from \SiIV\ $\lambda1393$ absorption in the range $-100<v_{\rm LSR}<100$\,\kms.}
\tn{b}{The log of the apparent column density ($N_a$ in cm$^{-2}$) measured from \SiIV\ $\lambda1402$ absorption in the range $-100<v_{\rm LSR}<100$\,\kms.}
\tn{c}{The adopted log apparent column density used for statistical analysis. See Section~\ref{section:observations} for a discussion on how $\mathrm{log}\,N_{a,{\rm adop.}}$ was calculated.}
\tn{c}{The residual apparent column density used for statistical analysis. See Section~\ref{section:Statistics} for a discussion on how $N_{a,R}$ was calculated.}
\tn{e}{Hubble Space Telescope Proposal ID number.}
\tn{f}{References: (1) \citet{Fox_2015} (2) \citet{Bordoloi_2017} (3) \citet{Karim_2018} (4) \citet{Ashley_2020}.}
\tn{g}{ND = not detected.}
\tn{h}{These measurements are affected by interfering emission (e.g. Lyman-$\alpha$ or intergalactic absorption) and were therefore set as an upper limit. If the line was saturated and affected by interfering absorption, then the line could not be measured and `...' was placed in the measurement.}
\tn{i}{HVC absorption identified in the associated reference is likely contributing to the LIV absorption in the $|v_{\rm LSR}|<100$ \kms.}
\end{deluxetable}

\section{Survival Analysis}\label{section:survival_example}
\begin{figure*}[!ht]
    \centering
    \epsscale{0.55} 
\plotone{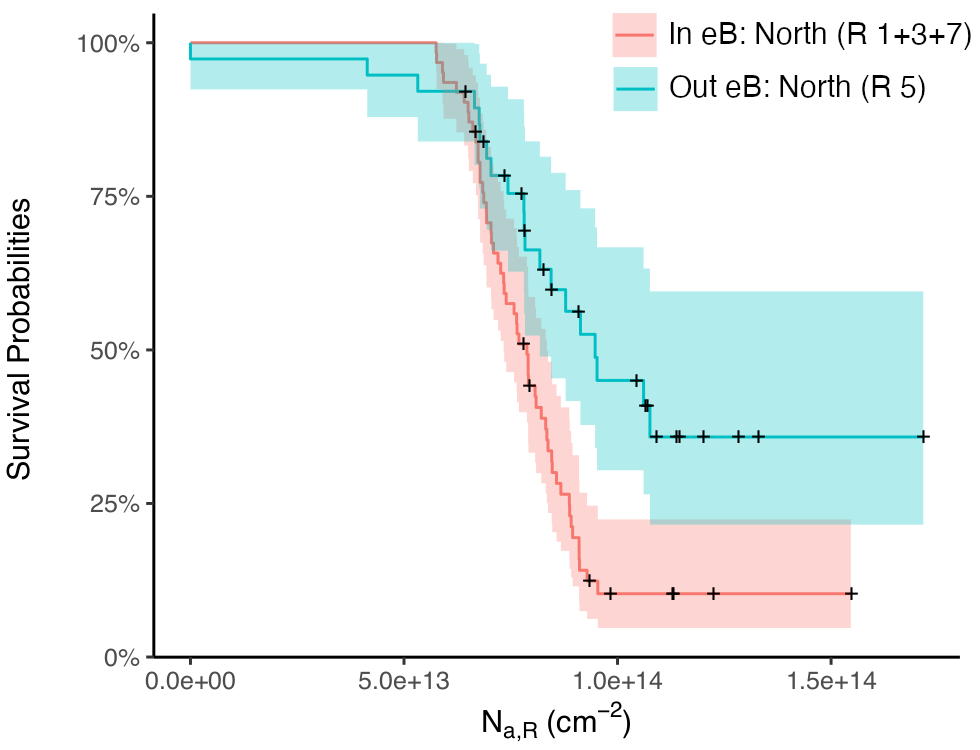}
\plotone{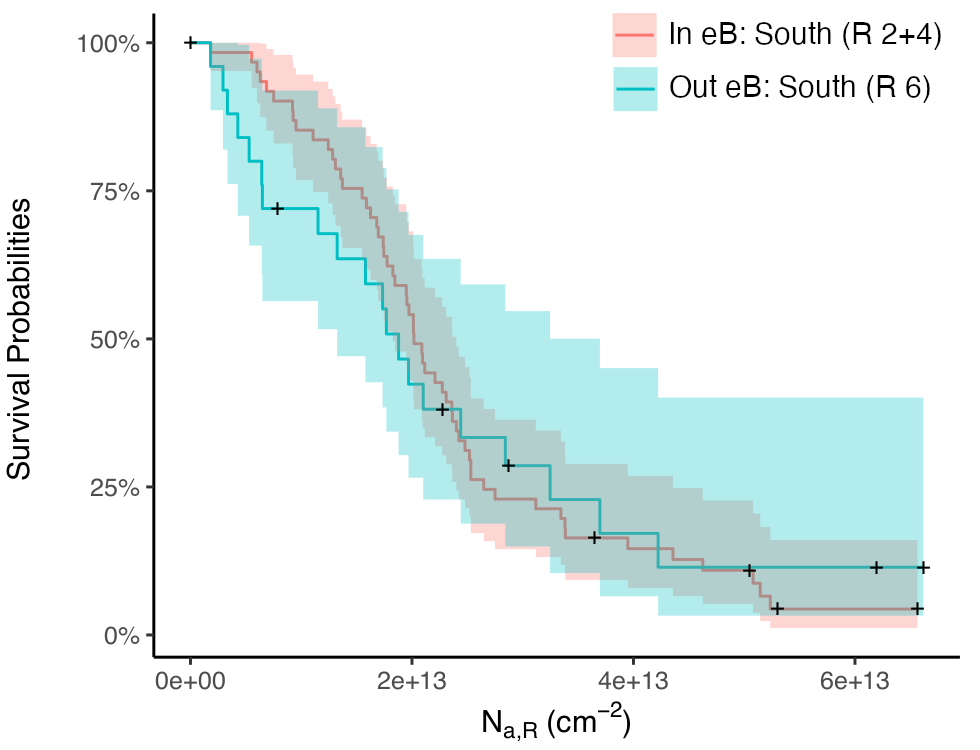}
    
\caption{Two examples of survival curves used in the analysis in Section~\ref{section:Statistics}. The residual columns displayed on the x-axes are offset by the minimum residuals for each pair of spatially compared groups,  $-7.6\times10^{13}$ and $-1.7\times10^{13}$ for the left and right plots, respectively, in order to obtain only positive numbers for the analysis. In both plots the shaded regions denote the 95\%\ confidence levels. Left: An example of different populations: inside the northern eB vs. outside the eB in the northern hemisphere. Right: An example of populations that were not shown to be different: inside the southern eB vs. outside the eB in the southern hemisphere.
\label{figure:survival_curves}}
\end{figure*}

In Figure~\ref{figure:survival_curves} we show two examples of survival curves created using \texttt{survfit} for the model-subtracted residual analysis. The plots start on the y-axis at 100\% ``survival'' and drop by the percent of sight lines with an unsaturated detection at the specified column density on the x-axis. The \texttt{survdiff} function determines if these two curves are drawn from different populations.

\end{document}